\documentclass[journal,twocolumn,table]{IEEEtran}

\usepackage{amsmath, amssymb, amsfonts, mathtools, amsthm, bbm, fdsymbol}

\usepackage{graphicx, adjustbox, subcaption, tikz, epsfig}

\usepackage{array, tabularx, booktabs, multirow, longtable, tabularray, supertabular, caption, makecell}

\usepackage{setspace, comment, float, rotating, hhline, soul, enumitem, textcase, graphicx}

\usepackage{cite, bibentry, bibunits}

\usepackage{xcolor}
\usepackage[hidelinks]{hyperref}
\hypersetup{hidelinks}

\usepackage[mathscr]{eucal}
\usepackage{threeparttable}
\usepackage{optidef}
\usepackage[noend]{algpseudocode} 
\usepackage[switch,pagewise]{lineno}
\usepackage[hmargin=1cm]{geometry}
\usepackage{nicematrix}
\usepackage{url, breakurl}

\usepackage{orcidlink}
\usepackage{array}
\usepackage{dblfloatfix}

\usepackage{multicol}

\usepackage{longtable}
\usepackage{pdflscape} % For landscape orientation
\usepackage{array}
\usepackage{geometry}
\usepackage{adjustbox} % For rotating content
\usepackage{amsmath}
\usepackage{geometry}
\usepackage{xr-hyper}
\usepackage[acronym]{glossaries}
\newacronym{cwnd}{W}{Congestion Window}
\newacronym{aimd}{AIMD}{Additive Increase Multiplicative Decrease}
\newacronym{tcp}{TCP}{Transmission Control Protocol}
\newacronym{rtt}{RTT}{Round Trip Time}
\newacronym{cca}{CCA}{Congestion Control Algorithm}
\newacronym{bbr}{BBR}{Bottleneck Bandwidth and Round-trip propagation time}
\newacronym{fcc}{FCC}{Federal Communications Commission}
\newacronym{bdp}{BDP}{Bandwidth-Delay Product}
\newacronym{dtc}{DTC}{Direct to Cell}
\newacronym{mbb}{MBB}{Make Before Break}
\newacronym{3gpp}{3GPP}{3rd Generation Partnership Project}
\newacronym{5g}{5G}{Fifth Generation}
\newacronym{6g}{6G}{6th-Generation}
\newacronym{ai}{AI}{Artificial Intelligence}
\newacronym{cnn}{CNN}{Convolutional Neural Network}
\newacronym{cogsat}{CogSat}{Cognitive Satellite}
\newacronym{cr}{CR}{Cognitive Radio}
\newacronym{crsn}{CRSN}{Cognitive Radio Sensor Networks}
\newacronym{csa}{CSA}{Concurrent Spectrum Access}
\newacronym{csi}{CSI}{Channel State Information}
\newacronym{dlr}{DLR}{Deep Learning Reinforcement}
\newacronym{dsa}{DSA}{Dynamic Spectrum Allocation}
\newacronym{dsrc}{DSRC}{Dedicated Short-Range Communications}
\newacronym{fl}{FL}{Federated Learning}
\newacronym{geo}{GEO}{Geostationary Equatorial Orbit}
\newacronym{iobt}{IoBT}{Internet of Battle Things}
\newacronym{iot}{IoT}{Internet of Things}
\newacronym{isl}{ISL}{Inter-Satellite Link}
\newacronym{itu}{ITU}{International Telecommunication Union}
\newacronym{leo}{LEO}{Low Earth Orbit}
\newacronym{lstm}{LSTM}{Long Short-Term Memory}
\newacronym{madrl}{MADRL}{Multi-Agent Deep Reinforcement Learning}
\newacronym{mdp}{MDP}{Markov Decision Process}
\newacronym{meo}{MEO}{Medium Earth Orbit}
\newacronym{ml}{ML}{Machine Learning}
\newacronym{nfv}{NFV}{Network Function Virtualization}
\newacronym{ntn}{NTN}{Non-Terrestrial Networks}
\newacronym{osa}{OSA}{Opportunistic Spectrum Access}
\newacronym{pu}{PU}{Primary User}
\newacronym{qos}{QoS}{Quality of Service}
\newacronym{rl}{RL}{Reinforcement Learning}
\newacronym{rsma}{RSMA}{Rate Splitting Multiple Access}
\newacronym{satcom}{SatCom}{Satellite Communication}
\newacronym{siot}{SIoT}{Satellite Internet of Things}
\newacronym{sl}{SL}{Supervised Learning}
\newacronym{su}{SU}{Secondary User}
\newacronym{uav}{UAV}{Unmanned Aerial Vehicle}
\newacronym{usl}{USL}{Unsupervised Learning}
\newacronym{vsat}{VSAT}{Very Small Aperture Terminal}
\newacronym{wsn}{WSN}{Wireless Sensor Networks}
\newacronym{gps}{GPS}{Global Positioning Systems}
\newacronym{dsm}{DSM}{Dynamic Spectrum Management}
\newacronym{stn}{STN}{Satellite Terrestrial Network}
\newacronym{eirp}{EIRP}{Effective Isotropic Radiated Power}
\newacronym{ss}{SS}{Sectrum Sensing}
\newacronym{kpi}{KPI}{Key Performance Indicator}
\newacronym{ca}{CA}{Channel Availability}
\newacronym{sinr}{SINR}{Signal to Noise plus Interference Ratio}
\newacronym{snr}{SNR}{Signal to Noise Ratio}
\newacronym{inr}{INR}{Interference to Noise Ratio}
\newacronym{ipc}{IPC}{Interference Power Constraint}
\newacronym{drl}{DRL}{Deep Reinforcement Learning}
\newacronym{ddpg}{DDPG}{Deep Deterministic Policy Gradient}
\newacronym{rem}{REM}{Radio Environment Map}
\newacronym{sdn}{SDN}{Software Defined Network}
\newacronym{rnn}{RNN}{Recurrent Neural Network}
\newacronym{dl}{DL}{Deep Learning}
\newacronym{svm}{SVM}{Support Vector Machine}
\newacronym{dqn}{DQN}{Deep Q-Network}
\newacronym{dca}{DCA}{Dynamic Channel Allocation}
\newacronym{wan}{WAN}{Wide Area Network}
\newacronym{gnss}{GNSS}{Global Navigation Satellite System}
\newacronym{ppo}{PPO}{Proximal Policy Optimization}
\newacronym{vits}{VITS}{Very High Throughput Satellites}
\newacronym{llm}{LLM}{Large Language Model}
\newacronym{sdr}{SDR}{Software Defined Radio}
\newacronym{css}{CSS}{Cooperative Spectrum Sensing}
\newacronym{ieee}{IEEE}{Institute of Electrical and Electronics Engineers}
\newacronym{rssi}{RSSI}{Received Signal Strength Indicator}
\newacronym{wran}{WRAN}{Wireless Regional Area Network}
\newacronym{cstn}{CSTN}{Cognitive Satellite Terrestrial Network}
\newacronym{bs}{BS}{Base Stations}
\newacronym{noma}{NOMA}{Non-Orthogonal Multiple Access}
\newacronym{cicstn}{CI-CSTN}{Cooperative Integrated-Cognitive Satellite Terrestrial Network}
\newacronym{hts}{HTS}{High Throughput Satellite}
\newacronym{mimo}{MIMO}{Multiple Input Multiple Output}
\newacronym{fss}{FSS}{Fixed Satellite Service}
\newacronym{mss}{MSS}{Mobile Satellite Service}
\newacronym{bss}{BSS}{Broadcasting Satellite Service}
\newacronym{mifr}{MIFR}{Master International Frequency Register}
\newacronym{wrc}{WRC}{World Radiocommunication Conference}
\newacronym{ngso}{NGSO}{Non-Geostationary Satellite Orbits}
\newacronym{mtc}{MTC}{Machine-Type Communications}
\newacronym{nbiot}{NB-IoT}{Narrowband Internet of Things}
\newacronym{etsi}{ETSI}{European Telecommunications Standards Institute}
\newacronym{bsm}{BSM}{Broadband Satellite Multimedia}
\newacronym{haps}{HAPS}{High-Altitude Platform Stations}
\newacronym{embb}{eMBB}{Enhanced Mobile Broadband}
\newacronym{mmtc}{mMTC}{Massive Machine Type Communications}
\newacronym{urllc}{URLLC}{Ultra-Reliable Low-Latency Communications}
\newacronym{ebu}{EBU}{European Broadcasting Union}
\newacronym{dth}{DTH}{Direct-to-Home}
\newacronym{acma}{ACMA}{Australian Communications and Media Authority}
\newacronym{cept}{CEPT}{European Conference of Postal and Telecommunications Administrations}
\newacronym{citel}{CITEL}{Inter-American Telecommunication Commission}
\newacronym{apt}{APT}{Asia-Pacific Telecommunity}
\newacronym{aws}{AWS}{Amazon Web Services}
\newacronym{esa}{ESA}{European Space Agency}
\newacronym{qpsk}{QPSK}{Quadrature Phase Shift Keying}
\newacronym{qam}{QAM}{Quadrature Amplitude Modulation}
\newacronym{tvws}{TVWS}{Television White Spaces}
\newacronym{darpa}{DARPA}{Defense Advanced Research Projects Agency}
\newacronym{iss}{ISS}{International Space Station}
\newacronym{scan}{SCaN}{Space Communications and Navigation}
\newacronym{knn}{KNN}{K-Nearest Neighbours}
\newacronym{pca}{PCA}{Principal Component Analysis}
\newacronym{gpt}{GPT}{Generative Pre-trained Transformer}
\newacronym{gan}{GAN}{Generative Adversarial Network}
\newacronym{fpga}{FPGA}{Field-Programmable Gate Array}
\newacronym{gpp}{GPP}{General-Purpose Processor}
\newacronym{vnf}{VNF}{Virtual Network Function}
\newacronym{mec}{MEC}{Multi Access Edge Computing}
\newacronym{usa}{USA}{United States of America}
\newacronym{uk}{UK}{United Kingdom}
\newacronym{vae}{VAE}{Variational Autoencoder}
\newacronym{los}{LoS}{Line of Sight}
\newacronym{bbm}{BBM}{Break Before Make}
\newacronym{acm}{ACM}{Adaptive Coding and Modulation}
\newacronym{ho}{HO}{Handover}
\newacronym{rb}{RB}{Resource Block}
\newacronym{prach}{PRACH}{Physical Random Access Channel}
\newacronym{isp}{ISP}{Internet Service Provider}
\newacronym{mcs}{MCS}{Modulation and Coding Scheme}
\newacronym{bbrv1}{BBR-v1}{BBR version 1}
\newacronym{bbrv2}{BBR-v2}{BBR version 2}
\newacronym{bbrv3}{BBR-v3}{BBR version 3}
\newacronym{tdm}{TDM}{Time Division Multiplexing}
\newacronym{pop}{PoP}{Point of Presence}

\makeglossaries

\begin{document}

\title{Unveiling TCP BBR Dominance in Starlink Internet: Experimental Insights and Analysis}

\author{
    Rakshitha De Silva \orcidlink{0000-0002-7194-7619}, 
    Shiva Raj Pokhrel \orcidlink{0000-0001-5819-765X}~\textit{{Senior~Member,~IEEE}} and
    Jonathan Kua \orcidlink{0000-0001-9699-9418}~\textit{{Member,~IEEE}}

    \thanks{This work is supported by SmartSat CRC, whose activities are funded by the Australian Government’s CRC Program.
    Authors are with the IoT \& Software Engineering Research Lab, School of Information Technology, Deakin University, Geelong, VIC 3125, Australia (e-mail: \href{mailto:rakshitha.desilva@deakin.edu.au}{rakshitha.desilva@deakin.edu.au}; \href{mailto:shiva.pokhrel@deakin.edu.au}{shiva.pokhrel@deakin.edu.au}; \href{mailto:jonathan.kua@deakin.edu.au}{jonathan.kua@deakin.edu.au}).\\
    }
}
\maketitle
 

\begin{abstract}
% This study provides an experimental assessment of Internet performance of Google's \gls{bbrv3} \gls{cca} over SpaceX's Starlink network using a six-city, five-continent testbed. We benchmark \gls{bbrv3} against eight \glspl{cca}---Cubic, Hybla, Vegas, LeoCC, Copa, PCC, \gls{bbrv1}, and \gls{bbrv2}---under dedicated and concurrent traffic scenarios. The results show that \gls{bbrv3}'s main strength is not aggressive bandwidth capture, but its balanced throughput, fairness, loss, and delay behavior over Starlink. We also develop practical models capturing Starlink path dynamics, including atmospheric attenuation, \glspl{isl}, and handover effects, to explain the observed performance trends. These findings provide useful information for the design of congestion control in next-generation \gls{leo} satellite Internet systems.
This experimental study delivers a global assessment of Google's Bottleneck Bandwidth and Round-trip propagation time-version 3 (BBR-v3) \gls{cca} over SpaceX's Starlink network.\color{black} 
Leveraging a strategically deployed six-city testbed across five continents, we systematically benchmark BBR-v3 against eight \glspl{cca}: Cubic, Hybla, Vegas, LeoCC, Copa, PCC, BBR-v1, and BBR-v2 under both dedicated and concurrent conditions. 
Our results demonstrate that BBR-v3's advantage is not aggressive bandwidth capture, but a more balanced fairness, loss, and delay trade-off over the Starlink Internet.
We develop pragmatic mathematical models that capture Starlink's complex network dynamics and characterize BBR-v3 behavior to better explain the experimental observations.
Our extensive evaluation of queue buildup and fairness further demonstrates BBR-v3's capability to maximize throughput in high-latency, variable satellite environments, while maintaining a balance between aggressiveness and fairness. 
The findings establish BBR-v3 as a compelling \gls{cca} for \gls{leo} satellite networks and provide a principled analytical foundation for next generation satellite Internet transport design.
\end{abstract}


\IEEEpeerreviewmaketitle

\vspace{-3mm}
%%%%%%%%%%%%%%%%%%%%% Section Introduction %%%%%%%%%%%%%%%%%%%%%%%%%%%%%%%%
\section{Introduction} 
\label{sec:Introduction}

\acrfull{leo} satellites have emerged as a cornerstone of next-generation global communications, providing wide-area broadband connectivity with significantly reduced latency compared to traditional \gls{geo} systems. Orbiting at altitudes below 2000 km, \gls{leo} mega-constellations bridge the digital divide by extending Internet access to underserved regions while offering competitive performance to terrestrial infrastructure. 
Among these, SpaceX's Starlink represents the most mature deployment.
Operating over 10,000 satellites across multiple orbital shells at $\approx$550 km altitude as of May 2026 and serving 2.7 million subscribers with 5 Tbps weekly capacity expansion~\cite{update-starlink, starlink_map}. 
Its architecture integrates phased-array Ku-band terminals, Ka-band feeder links, and laser \acrfullpl{isl}, enabling global connectivity even in gateway-sparse regions and transforming \gls{leo} constellations from supplemental backhaul into global \glspl{isp}.

\begin{figure}[!t]
\centering
    \includegraphics[width=\linewidth]{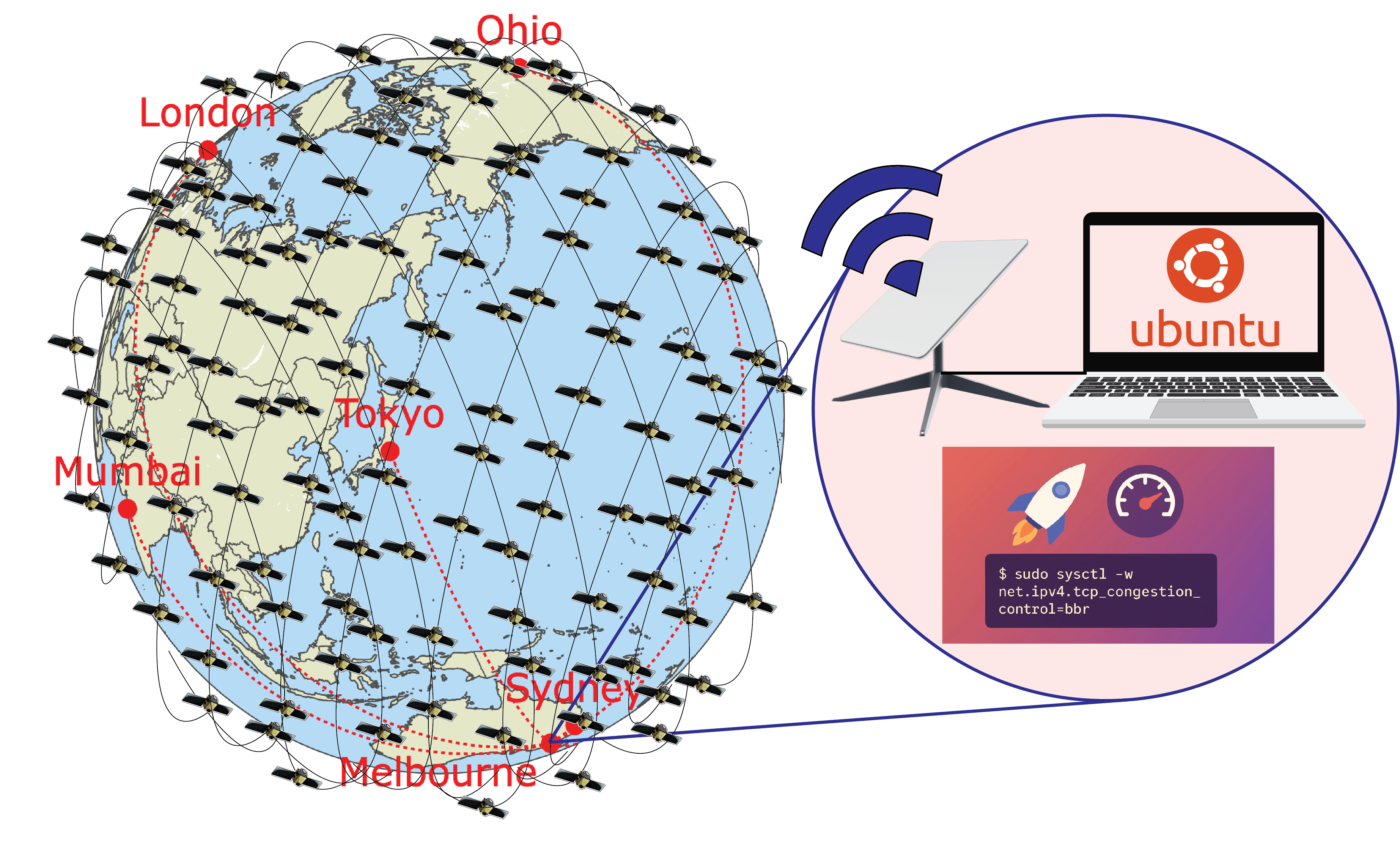}
    \vspace{-7mm}
  \caption{A simplified illustration of our global experimental testbed over Starlink Internet.}
\label{fig:testbed_globe}
\end{figure}

Google's \gls{bbr}, introduced in 2017, represents a paradigm shift in \gls{tcp} \acrfullpl{cca} by explicitly modeling bottleneck bandwidth and propagation delay rather than relying on loss-based congestion detection. While \gls{bbrv1} demonstrated significant throughput improvements, it revealed fairness and efficiency challenges, particularly persistent queue buildup. \gls{bbrv2} addressed these issues but continued to struggle in heterogeneous environments~\cite{scherrer2022model, bbr_survey}. The latest iteration, \gls{bbrv3}, refines probing and pacing strategies to optimize throughput, latency, and fairness across diverse network conditions, with ongoing research characterizing its behavior in various deployment scenarios~\cite{bbrv3}.

The convergence of Starlink's global \gls{leo} network with \gls{bbr}'s model-based congestion control presents significant yet underexplored potential. Existing research on \gls{bbr} over \gls{leo} networks primarily relies on simulations~\cite{barbosa2023comparative} or single-connection analyses predating \gls{bbrv3}~\cite{claypool2021comparison, mobility2024yang}. Studies using actual Starlink data have focused on video streaming application-level performance~\cite{mohan2024multifaceted, izhikevich2024global}, rather than \gls{bbr}'s network-level performance. 
Furthermore, most work on \gls{tcp} over \gls{leo} research does not cover physical layer evaluation, critical to transport layer performance.
\color{black}
Addressing these gaps, this work makes the following contributions:
\begin{itemize}
\item A decomposition of the Starlink network into user link, \glspl{isl}, and feeder link components with mathematical models yielding end-to-end transmission path failure probability.
\item Design and implementation of a global Starlink Internet performance testbed across six cities: Ohio, São Paulo, London, Mumbai, Tokyo, and Sydney.
\item Extensive analyses of \gls{bbrv3} performance comparison against eight \glspl{cca}: Cubic \cite{cubic_rfc}, Vegas \cite{low2002understanding}, Hybla~\cite{caini2004tcp}, LEO Network Congestion Control (LeoCC)~\cite{lai2025leocc}, Copa~\cite{arun2018copa}, Performance-oriented Congestion Control (PCC)~\cite{dong2015pcc}, \gls{bbrv1}, and \gls{bbrv2}. The evaluation covers four primary network conditions: dedicated uplink and downlink, inter-\gls{cca} concurrent uplink and downlink. 
\item Present a fluid model of \gls{bbrv3} congestion window modeling yielding end-to-end transmission path and transmission failure probability that capture Starlink LEO dynamics and explain experimental observations.
\item Queuing analysis and fairness evaluation of \gls{bbrv3} encapsulating the captured data over the globally distributed testbed.
\end{itemize}
\color{black}

The remainder of this paper is organized as follows: Section~\ref{sec:starlink network} provides an overview of the Starlink network characteristics and derives a mathematical model of the end-to-end link failure probability.
Section \ref{sec:testbed_results} details the globally distributed testbed implementations and provides an evaluation of the collected results. 
Section \ref{sec:fluid_model} presents the fluid model, queuing analysis, and provides a fairness evaluation of \gls{bbrv3} encapsulating the collected data, and Section \ref{sec:Conclusions} concludes the article.

\section{Starlink Internet: System and Settings}
\label{sec:starlink network}

\subsection{Overview of the Starlink Network}
\label{subsec:starlink_overview}

As of May 2026, there are 10,408 active Starlink satellites in orbit, making SpaceX's mega-\gls{leo} constellation by far the largest of its kind \cite{starlink_map}.
% SpaceX's Starlink official data indicates that as of February 2025, the LEO constellation comprises 6,751 operational satellites from the 7,946 launched. 
These satellites are categorized into four variants: V1, V1.5, V2-KU (utilizing Ku-band), and V2-DTC, a specialized version of \gls{dtc} designed for direct smartphone connectivity \cite{update-starlink}. 
According to \gls{fcc} filings, the constellation is structured into five distinct orbital shells with different orbital lanes and inclination angles \cite{pachler2021updated}.
% characteristics detailed in Table \ref{tab:shells} \cite{pachler2021updated}. 
This architecture ensures high satellite availability, with over 20 satellites simultaneously visible in densely populated mid-latitude regions upon full deployment \cite{del2019technical}. 
Notably, Starlink deviates from the traditional Walker constellation designed to facilitate seamless coverage through its multi-shell architecture while maintaining consistent ground tracks \cite{jing2024analysis}.

% \begin{table}[t]
%     \centering
%     \caption{Different shells and their characteristics of Starlink}
%     \resizebox{\linewidth}{!}{
%     \begin{tabular}{c|c|c|c|c}
%         \hline
%         \textbf{Shell} & \textbf{Altitude (km)} & \textbf{Orbital planes} & \textbf{LEOs per plane} & \textbf{Inclination} \\ \hline
%         1 & 540 & 72 & 22 & $53.2^\circ$ \\ \hline
%         2 & 550 & 72 & 22 & $53^\circ$ \\ \hline
%         3 & 560 & 6 & 58 & $97.6^\circ$ \\ \hline
%         4 & 560 & 4 & 43 & $97.6^\circ$ \\ \hline
%         5 & 570 & 56 & 20 & $70^\circ$ \\ \hline
%     \end{tabular}
%     }
%     \label{tab:shells}
% \end{table}
% \vspace{-5mm}

\begin{figure}[!t]
\centering
    \includegraphics[width=\linewidth]{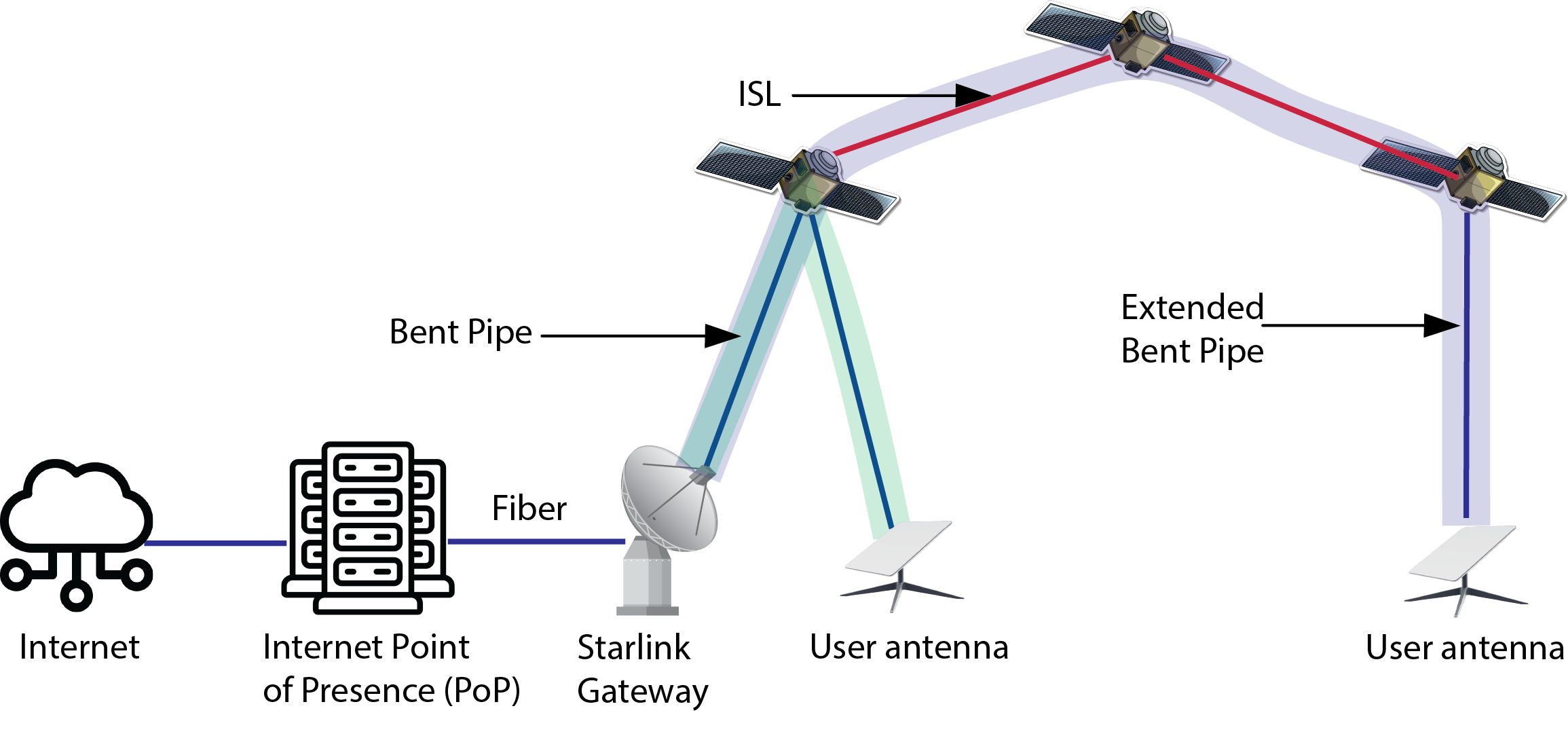}
    \vspace{-7mm}
  \caption{\textcolor{black}{Traffic flow through ``bent-pipes'' in the Starlink network.}}
\label{fig:bentpipe}
\end{figure}

Starlink's network topology leverages optical \glspl{isl} (colloquially ``space lasers") that enable service provision without requiring gateway presence in the satellite's coverage area. 
Each satellite incorporates three 200 Gbps optical \glspl{isl}, collectively forming a global mesh network \cite{update-starlink}. 
This architecture significantly enhances coverage over oceanic and remote regions while reducing ground station infrastructure requirements. 
The globally distributed gateway network operates in Ka-band, with each gateway capable of simultaneously connecting to four satellites. 
Gateway downlinks utilize nine 250 MHz channels (17.8-19.3 GHz), while uplinks employ eight 500 MHz channels (27.5-30.0 GHz) \cite{del2019technical}.
User terminals (``Dishy") employ Ku-band transmission, connecting to \glspl{leo} visible above $25^\circ$ elevation angle. 
Through multi-beam antenna technology, a single satellite can simultaneously service multiple users. 
As illustrated in Fig. \ref{fig:bentpipe}, user data traverses a "bent-pipe" connection across \glspl{isl} \cite{mohan2024multifaceted,alcs2025wang}. 
The \gls{acma} has confirmed Starlink's \gls{itu} frequency allocations in Australia spanning 10.7–12.7 GHz, 14–14.5 GHz, 17.8–18.55 GHz, 18.8–19.3 GHz, 27.5–29.1 GHz, and 29.5–30 GHz \cite{acma-starlink}.

\begin{figure}[tb]
    \centering

    \begin{subfigure}{\columnwidth}
        \centering
        \includegraphics[width=\linewidth]{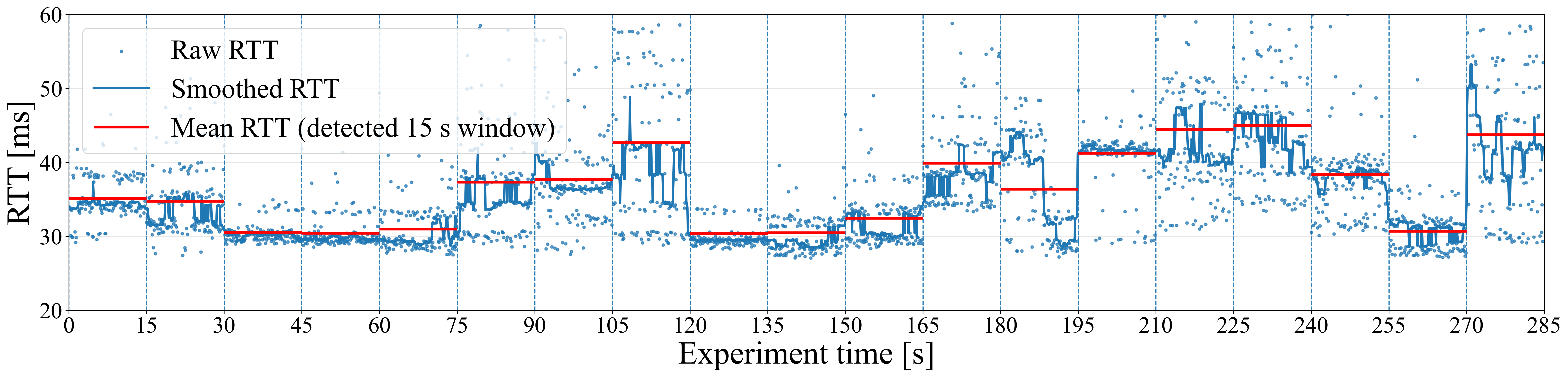}
        \caption{Sydney to Melbourne}
        \label{fig:sub1}
    \end{subfigure}

    \vspace{2mm}

    \begin{subfigure}{\columnwidth}
        \centering
        \includegraphics[width=\linewidth]{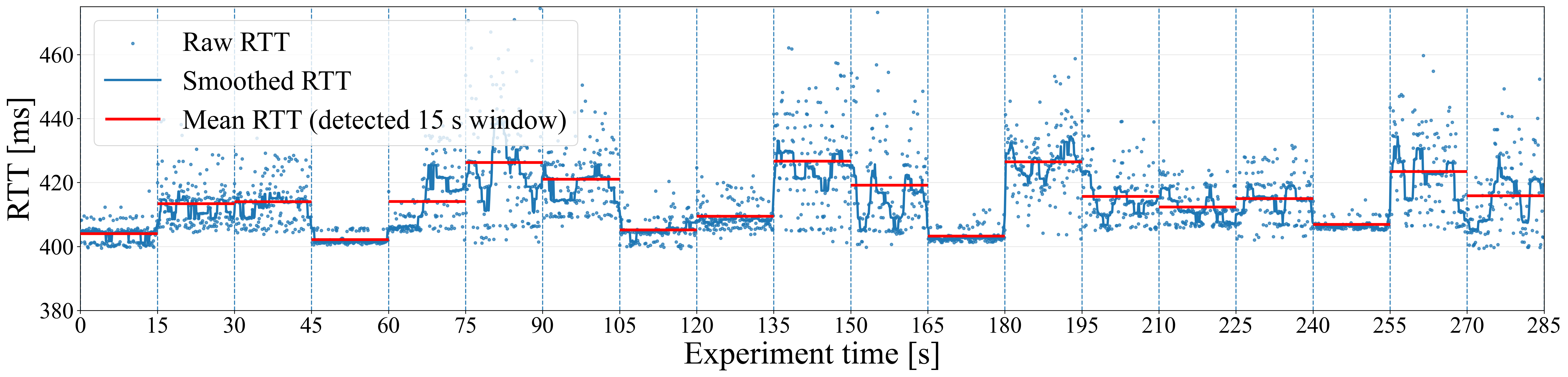}
        \caption{São~Paulo to Melbourne}
        \label{fig:sub2}
    \end{subfigure}

    \caption{RTT shift at 15~s boundaries in Starlink network}
    \label{fig:two_subplots_vertical}
\end{figure}

The ground terminals leverage phased-array antennas, comprising a grid of smaller antennas that enable electronic beam steering through differential phase manipulation. 
Each terminal contains five Ku-band phased array antennas and three dual-band antennas (Ku and E bands) for user connectivity. 
To counteract \gls{snr} variations caused by orbital dynamics, handovers, and environmental factors, Starlink dynamically adjusts modulation and coding schemes \cite{humphreys2023signal}. 
User downlinks operate across eight 250 MHz channels (each with 10 MHz guard bands), while uplinks utilize four 125 MHz channels \cite{humphreys2023signal, del2019technical}. 
The transmission structure employs time division multiplexing with a 1/750 s frame period subdivided into 4.4 $\mu$s frames and guard intervals, with frame headers containing satellite, channel, and modulation information \cite{humphreys2023signal}.
\color{black}
The \gls{rtt} traces captured using \texttt{irtt} for two internet paths presented in Fig.~\ref{fig:two_subplots_vertical} exhibit a unique characteristic of the Starlink constellation. 
The repeated step-like \gls{rtt} shifts are experienced close to 15~s boundaries, where each interval tends to remain within a relatively stable regime before transitioning to a different level in the subsequent window. 
This behavior is a result of planned ``connection hand-offs'' between the Dishy and the \gls{leo} satellite to maintain a seamless connectivity amid the constellation dynamics, which is further detailed in Starlink's FCC filing \cite{starlink_fcc}.
\color{black}

\subsection{End-to-End Link Failure and Packet Drop Probability}
\label{subsec:e2e_prob}

As illustrated in Fig.~\ref{fig:bentpipe}, an end-to-end Starlink transmission path may consist of three main wireless components subject to uncertainty: the gateway-to-\gls{leo} link, \glspl{isl}, and the \gls{leo}-to-ground-terminal link. 
The achievable end-to-end capacity can therefore be affected by propagation latency variations due to orbital dynamics, \gls{acm}-driven changes in response to time-varying \gls{snr}, satellite handovers, and bandwidth contention across shared \glspl{isl} \cite{al2023deep, khan2022rate, weththasinghe2024optimising}.
Accounting for these factors and the distinct transmission legs of the Starlink link, we present an end-to-end link failure and packet drop probability model.  
The model encapsulates four key impairment terms, the packet drop probability due to \gls{leo} satellite to ground terminal capacity limitations ($p^{cap}$) \cite{song2024analysis, latency2023pan}, the failure probability due to atmospheric attenuation ($p^{at}$) \cite{ITU_AtmosphericLoss2019, ITU_R_P.676_13}, the handover drop probability ($p^{ho}$) \cite{hreha2019synchronization, starlink_fcc}, and the \gls{isl} failure probability ($p_i^{\text{isl}}$) \cite{zhu2022laser}. 
The detailed analysis is provided under Appendix~\ref{appendix:link_analysis}.

\color{black}
The probability of link failure causing packet loss in the transmission path ($p^{\text{tot}}$) can be given as:
\begin{equation}
    p^{\text{tot}} = 1 - (1-p^{cap}) \cdot(1-p^{gw}) \cdot (1-p^{at})\cdot (1-p^{ho}) \cdot \prod_{i=0}^{N^{\text{isl}}} (1-p_i^{\text{isl}} )
    \label{eq:total_prop1}
\end{equation}
where $N^{\text{isl}}$ is the number of \glspl{isl} that a given packet is routed through in the constellation, $p^{gw} = p^{at}(gw)$, and the other parameters represent the respective failure probabilities~\cite{latency2023pan}.
For simplicity, we assume that gateway links are negligibly affected by handover losses.
If the number of \glspl{isl} in a given extended bent pipe over the Starlink network ($N^{\text{isl}}$) is deterministic, Eq. \eqref{eq:total_prop1} becomes: 
\begin{equation}
p^{\text{tot}} = 1- (1-p^{cap}) \cdot(1-p^{gw}) \cdot (1-p^{at})\cdot (1-p^{ho}) \cdot (1-p^{\text{isl}})^{N^{\text{isl}}} 
\label{eq:total_prob}
\end{equation}
\color{black}

\section{Testbed Setup and CCA Performance Over Starlink}
\label{sec:testbed_results}

This section presents our globally distributed testbed implementation across six test sites and compares the performance of \gls{bbrv3} against eight CCAs: Cubic \cite{cubic_rfc}, Vegas \cite{low2002understanding}, Hybla~\cite{caini2004tcp}, LeoCC~\cite{lai2025leocc}, Copa~\cite{arun2018copa}, PCC~\cite{dong2015pcc}, \gls{bbrv1}, and \gls{bbrv2} using it over the Starlink network.

\subsection{Testbed Setup}
\label{subsec:testbed}

\begin{figure}[t]
\centering
    \includegraphics[width=0.8\linewidth]{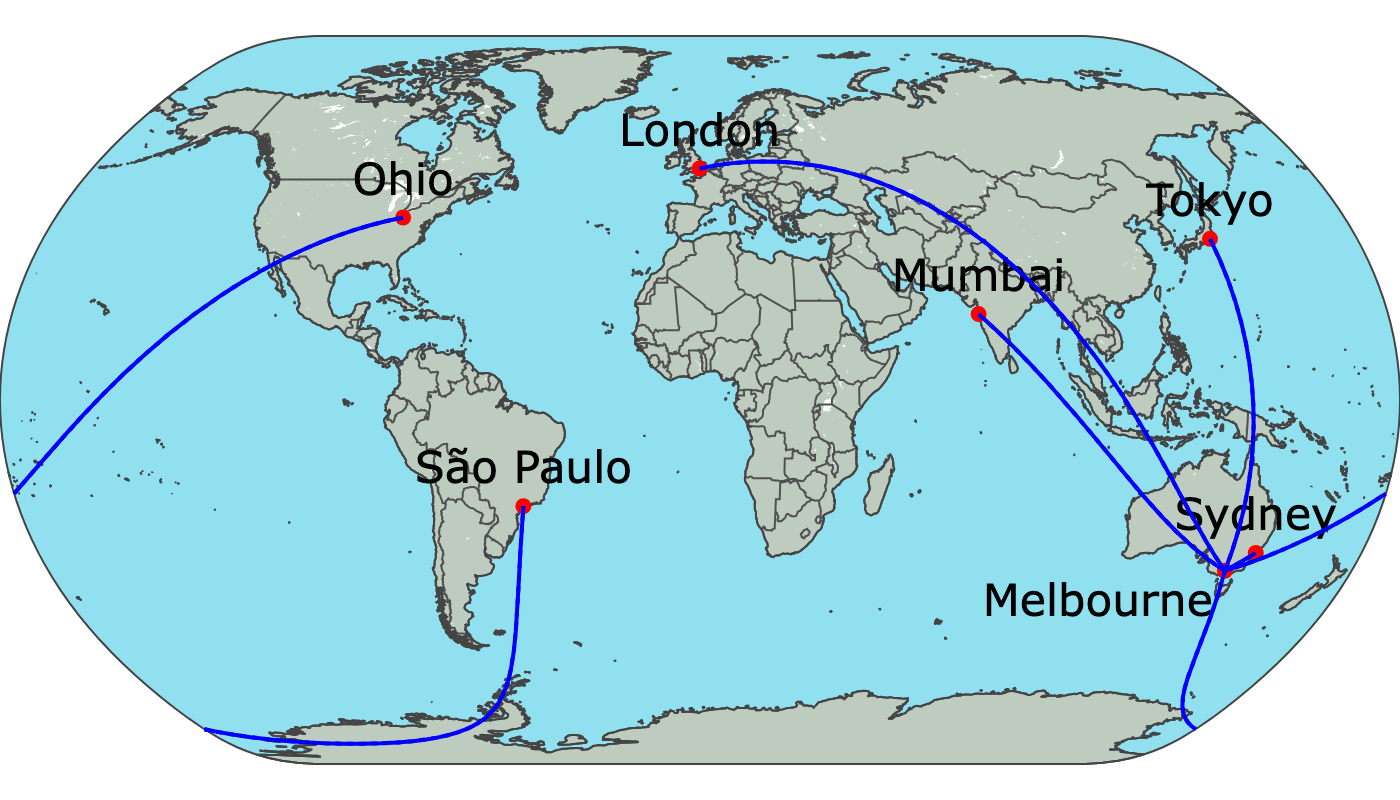}
    \vspace{-2mm}
  \caption{Globally distributed server locations}
\label{fig:map}
\end{figure}

\color{black}
As illustrated in Fig. \ref{fig:map}, we set up Linux servers in six main cities distributed globally, namely Ohio, São Paulo, London, Mumbai, Tokyo, and Sydney. 
We leveraged \gls{aws} Elastic Compute Cloud~(EC2) for this distributed server setup, and the local portion was hosted on the university premises in Burwood, Melbourne, Australia. 
\gls{aws} EC2 instances were set up with \texttt{Ubuntu~24.04.4~LTS} Linux distribution, and the default free-tier settings were used, apart from the necessary port security group configurations. 
Furthermore, we assume the \gls{aws} cloud instances to provide a consistent network connection throughout the geographically distributed cloud instances. 
The Starlink user terminal consists of the latest-generation standard kit with a UTA-232 model dish. 
The local Linux server (\texttt{Ubuntu~24.04.4~LTS}) was connected to the Starlink terminal via a Category 6 Ethernet connection, creating a continuous test environment. 
To test \gls{tcp} \gls{cca} performance over the testbed, we leverage \texttt{iperf3}, an open-source network testing tool used to measure the maximum achievable bandwidth and performance over network connections.
It provides detailed metrics such as throughput, packet loss, jitter, and retransmissions, making it widely used for network diagnostics and benchmarking. 
To identify the Starlink \gls{pop} associated with our connection, we performed repeated logging and DNS resolving during active flows to multiple AWS regions, and the reverse DNS consistently resolved to \texttt{customer.mlbeaus1.isp.starlink.com}, indicating a stable Starlink \gls{pop} association in the Melbourne region during the tests rather than destination-dependent \gls{pop} switching.

We implemented an automated measurement framework using \texttt{iperf3}, in which each remote EC2 server and the local host were configured with the required \gls{cca}. 
We define tests with a single active \gls{tcp} flow as the \emph{dedicated} flow scenario, in which forward and reverse \texttt{iperf3} tests were conducted to collect dedicated downlink and uplink measurements for each \gls{cca}.
To evaluate inter-\gls{cca} coexistence under shared-path contention, we define the \emph{concurrent} flow scenario. 
In this test, multiple \gls{tcp} flows run simultaneously over the same Starlink path, with each flow configured to use a different \gls{cca}. 
Under \emph{concurrent} flows, nine simultaneous \gls{tcp} flows were generated, corresponding to Cubic, Hybla, Vegas, LeoCC, Copa, PCC, \gls{bbrv1}, \gls{bbrv2}, and \gls{bbrv3}. 
Each flow was isolated in a separate Linux network namespace, allowing its corresponding \gls{cca} to be configured independently while sharing the same end-to-end Starlink path.
Special care was taken to synchronize server startup, ensuring data flushing before retrieval, and preserve namespace-level isolation across \glspl{cca}. 
This enabled fair and repeatable benchmarking under Starlink network conditions. 
For each \gls{cca}, destination, direction, and scenario combination, we collected 10 independent runs, each with a 300-second capture window. The full measurement campaign was carried out during April 2026.
\color{black}

\color{black}
\subsection{Evaluation of Downlink Streams}
\label{subsec:download}

\begin{figure*}[h!]
\centering
    \includegraphics[width=\textwidth]{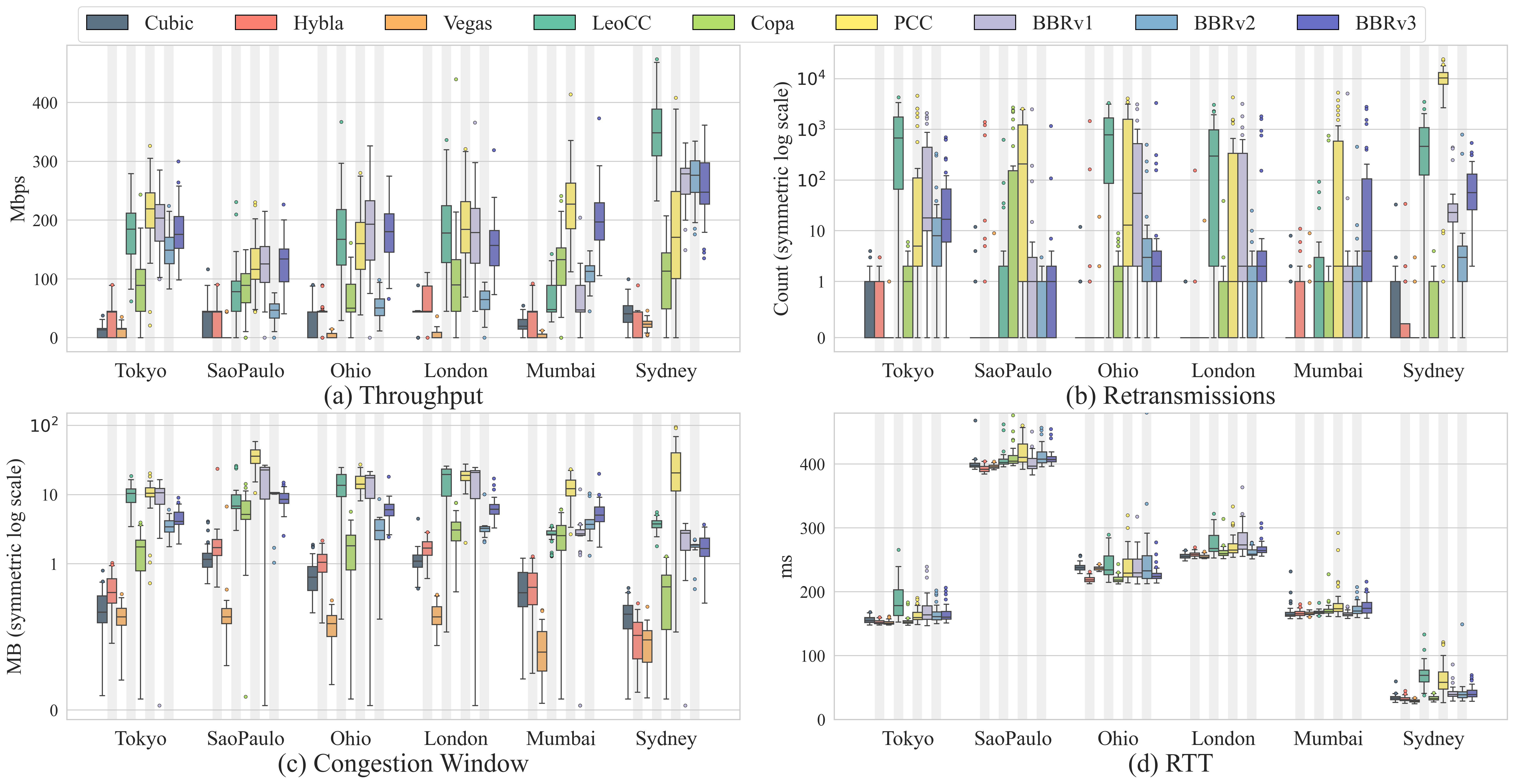}
  \caption{Summarized downlink observations over Starlink with dedicated \glspl{cca} for globally distributed locations}
\label{fig.download box}
\end{figure*}

\begin{figure*}[h!]
\centering
    \includegraphics[width=\textwidth]{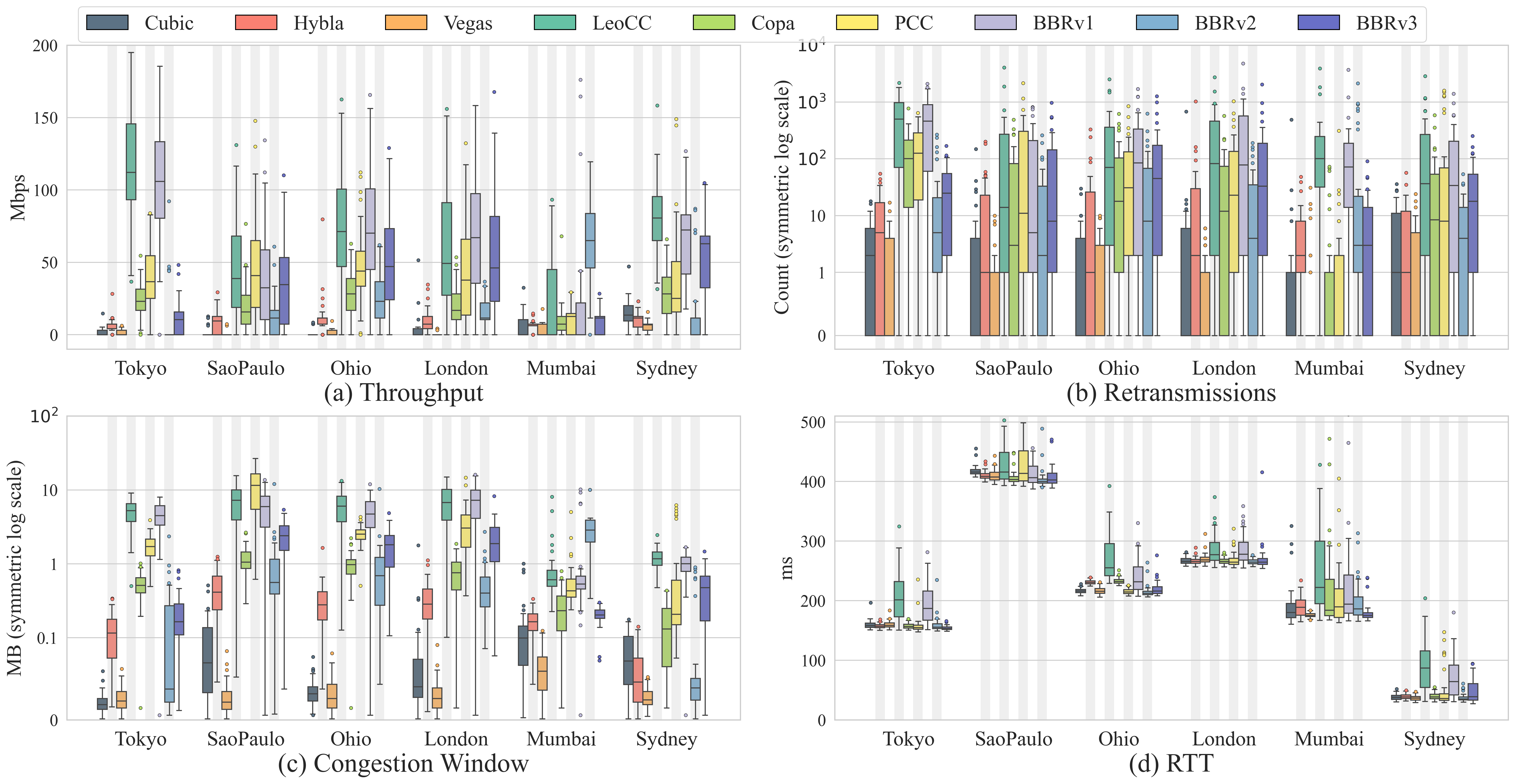}
    % \vspace{-10mm}
  \caption{Summarized downlink observations over Starlink with concurrent \glspl{cca} for globally distributed locations}
\label{fig.download competitive box}
\end{figure*}

\subsubsection{Dedicated Downlink Streams}
\label{subsubsec:download_individual}

The dedicated-flow downlink results in Fig.~\ref{fig.download box} show that \gls{bbrv3} provides a strong but not universally dominant operating point over Starlink. 
As shown in Fig.~\ref{fig.download box}(a), \gls{bbrv3} achieves consistently high median throughput across all six locations, with values of 176.11~Mbps in Tokyo, 134.11~Mbps in São~Paulo, 180.37~Mbps in Ohio, 157.14~Mbps in London, 197.21~Mbps in Mumbai, and 247.56~Mbps in Sydney. 
This is substantially higher than Cubic, Hybla, Vegas, and Copa in most locations, whose rates remain limited or highly variable. 
However, LeoCC, PCC, \gls{bbrv1}, and \gls{bbrv2} exceed \gls{bbrv3} in some paths, indicating that more aggressive and LEO-aware schemes can capture additional capacity under favorable Starlink conditions. 
This higher capacity capture is accompanied by higher retransmission counts and larger congestion windows, as presented in Fig.~\ref{fig.download box}(b) and Fig.~\ref{fig.download box}(c).
Interestingly, \gls{bbrv3} maintains a more controlled congestion-window profile while avoiding severe retransmission penalties. 
The \gls{rtt} distributions in Fig.~\ref{fig.download box}(d) reflect the geographic path differences across the six endpoints, and highlight \gls{bbrv3}'s \gls{rtt} stability over Starlink Internet. 
The receiver-advertised window and RTT-variance associated with these captures are presented in Fig.~\ref{fig.download_seq_fig2_rwnd_rttvar}, and the summary of the data captured from 10 dedicated download tests is presented in Fig.~\ref{fig:downlink_10_runs} under Appendix~\ref{appendixC}. 
Those results further indicate that \gls{bbrv3} offers a favorable throughput, loss, and delay trade-off behavior over the Starlink Internet in comparison to the evaluated \glspl{cca}.

\subsubsection{Concurrent Downlink Streams}
\label{subsubsec:download_parallel}

The concurrent downlink results in Fig.~\ref{fig.download competitive box} show a more contested operating regime than the dedicated downlink case, as all nine \glspl{cca} simultaneously share the same Starlink path. 
In this setting, \gls{bbrv3} remains competitive but does not dominate the bandwidth allocation. 
As shown in Fig.~\ref{fig.download competitive box}(a), \gls{bbrv3} records median throughput values of 10.49~Mbps in Tokyo, 34.58~Mbps in São~Paulo, 47.17~Mbps in Ohio, 46.11~Mbps in London, 11.54~Mbps in Mumbai, and 62.87~Mbps in Sydney. 
These values are generally higher than the conservative Cubic, Hybla, Vegas, and Copa flows, but lower than LeoCC and \gls{bbrv1}, indicating that more aggressive or LEO-aware schemes capture a larger share of the available downlink capacity under concurrent contention.
This behavior is reflected in Fig.~\ref{fig.download competitive box}(c), where LeoCC, \gls{bbrv1}, and PCC often maintain larger congestion windows, while \gls{bbrv3} operates with a more restrained in-flight volume. 
The retransmission results in Fig.~\ref{fig.download competitive box}(b) show the cost of this aggressive capacity capture: LeoCC and \gls{bbrv1} incur substantially higher retransmission counts in several paths, especially Tokyo, whereas \gls{bbrv3} maintains moderate retransmissions and avoids a severe loss behavior. 
The \gls{rtt} distributions in Fig.~\ref{fig.download competitive box}(d) continue to reflect the geographic path differences, while also highlighting that aggressive \glspl{cca} can result in increased delay. 
The receiver-advertised window and \gls{rtt}-variance results for the same concurrent downlink captures are provided in Fig.~\ref{fig.download_competitive_fig2_rwnd_rttvar} in Appendix~\ref{appendixC}, together with the 10-run summary in Fig.~\ref{fig:downlink_10_runs}. 
The results indicate that \gls{bbrv3} sacrifices some bandwidth share compared with the most aggressive contenders, but provides a more controlled loss and delay profile under heterogeneous \gls{cca} competition.

\subsection{Evaluation of Uplink Streams}
\label{subsec:upload}

\begin{figure*}[h!]
\centering
    \includegraphics[width=\textwidth]{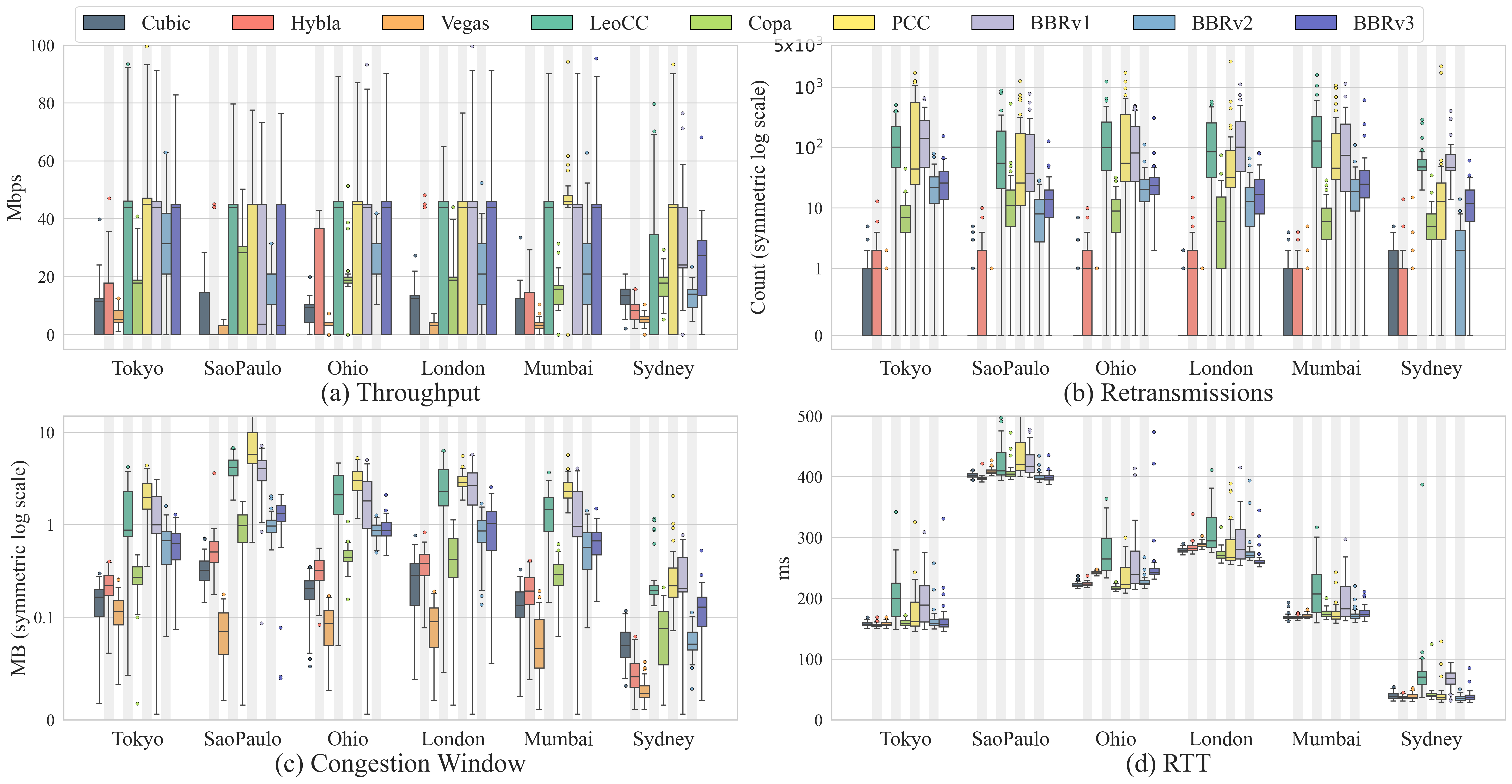}
    \vspace{-5mm}
  \caption{Summarized uplink observations over Starlink with dedicated \glspl{cca} for globally distributed locations}
\label{fig.upload_box}
\end{figure*}

\begin{figure*}[h!]
\centering
    \includegraphics[width=\textwidth]{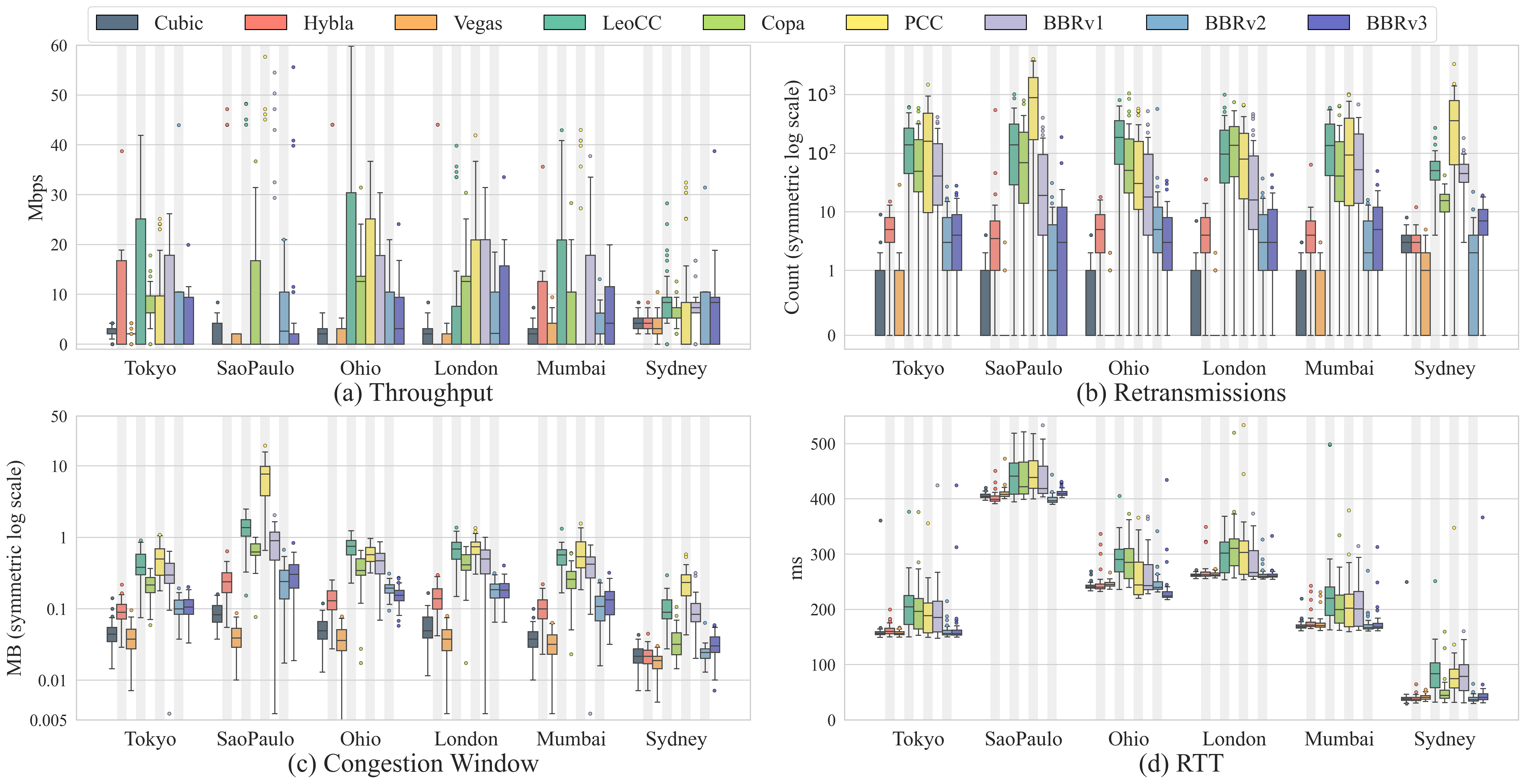}
    \vspace{-5mm}
  \caption{Summarized downlink observations over Starlink with concurrent \glspl{cca} for globally distributed locations}
\label{fig.parallel upload box}
\end{figure*}

\subsubsection{Dedicated Uplink Streams}
\label{subsubsec:upload_individual}

The dedicated-flow uplink results in Fig.~\ref{fig.upload_box} show that the Starlink uplink presents a more capacity-constrained operating regime than the downlink case. 
This is primarily due to the limited bandwidth availability on the Starlink uplink. 
As shown in Fig.~\ref{fig.upload_box}(a), \gls{bbrv3} achieves strong median uplink throughput in Tokyo, Ohio, London, and Mumbai, with values close to 44~Mbps, while recording a lower median, but a higher lower quartile in Sydney. 
This indicates that \gls{bbrv3} can efficiently utilize the available uplink capacity when the path is stable, but its performance remains sensitive to path-specific Starlink Internet conditions. 
LeoCC, PCC, and \gls{bbrv1} achieve comparable or higher throughput in several locations, but this is accompanied by larger congestion windows and higher retransmission counts, as shown in Fig.~\ref{fig.upload_box}(c) and Fig.~\ref{fig.upload_box}(b), respectively. 
In contrast, Cubic, Hybla, Vegas, and Copa generally remain more conservative, producing lower or less stable uplink throughput. 
The \gls{rtt} results in Fig.~\ref{fig.upload_box}(d) again reflect the geographic path differences, with Sydney showing the lowest delay and São~Paulo the highest. 
These dedicated-flow uplink results show that \gls{bbrv3} provides a competitive uplink operating point, balancing throughput with more controlled retransmission and congestion-window behavior compared with the most aggressive \glspl{cca}. 
The receiver-advertised window and \gls{rtt}-variance results associated with these captures are provided in Fig.~\ref{fig.uplink_seq_fig2_rwnd_rttvar}, and the 10-run summary is presented in Fig.~\ref{fig:uplink_10_runs} under Appendix~\ref{appendixC}.

\subsubsection{Concurrent Streams}
\label{subsubsec:upload_parallel}

The concurrent uplink results in Fig.~\ref{fig.parallel upload box} show that simultaneous CCA competition over the Starlink uplink creates a strongly capacity-constrained regime, where most algorithms experience intermittent or low median throughput. 
As shown in Fig.~\ref{fig.parallel upload box}(a), \gls{bbrv3} achieves non-zero median throughput in Tokyo, Ohio, Mumbai, and Sydney, but drops to near-zero median throughput in São~Paulo and London.
Underlining that its uplink performance under contention is sensitive to path-specific capacity variation and competing-flow pressure. 
\gls{bbrv2} and LeoCC achieve higher medians in selected locations, while Cubic, Vegas, Copa, and PCC remain limited or highly bursty in most cases. 
The retransmission and congestion-window results in Fig.~\ref{fig.parallel upload box}(b) and Fig.~\ref{fig.parallel upload box}(c) show that the algorithms that capture more uplink capacity often do so with higher retransmission counts or larger in-flight volumes.
Whereas \gls{bbrv3} generally maintains a more restrained congestion-window profile and avoids the most severe loss behavior. 
The \gls{rtt} distributions in Fig.~\ref{fig.parallel upload box}(d) mainly follow the geographic path differences, while also reflecting additional delay variation introduced to aggressive \glspl{cca} by simultaneous uplink contention. 
The receiver-advertised window and \gls{rtt}-variance associated with these concurrent uplink captures are presented in Fig.~\ref{fig.uplink_competitive_fig2_rwnd_rttvar}, and the 10-run summary is provided in Fig.~\ref{fig:uplink_10_runs} under Appendix~\ref{appendixC}. 
To this end, the concurrent uplink results further highlight the fact that \gls{bbrv3} does not consistently dominate over Starlink Internet, but provides a controlled operating point with moderate throughput, restrained congestion-window growth, and lower loss exposure than the most aggressive competing alternatives.

\subsection{TCP CCA Operating Regimes}
\label{subsec:CCA_operations}

\begin{figure}[tb]
\centering
    \includegraphics[width=\linewidth]{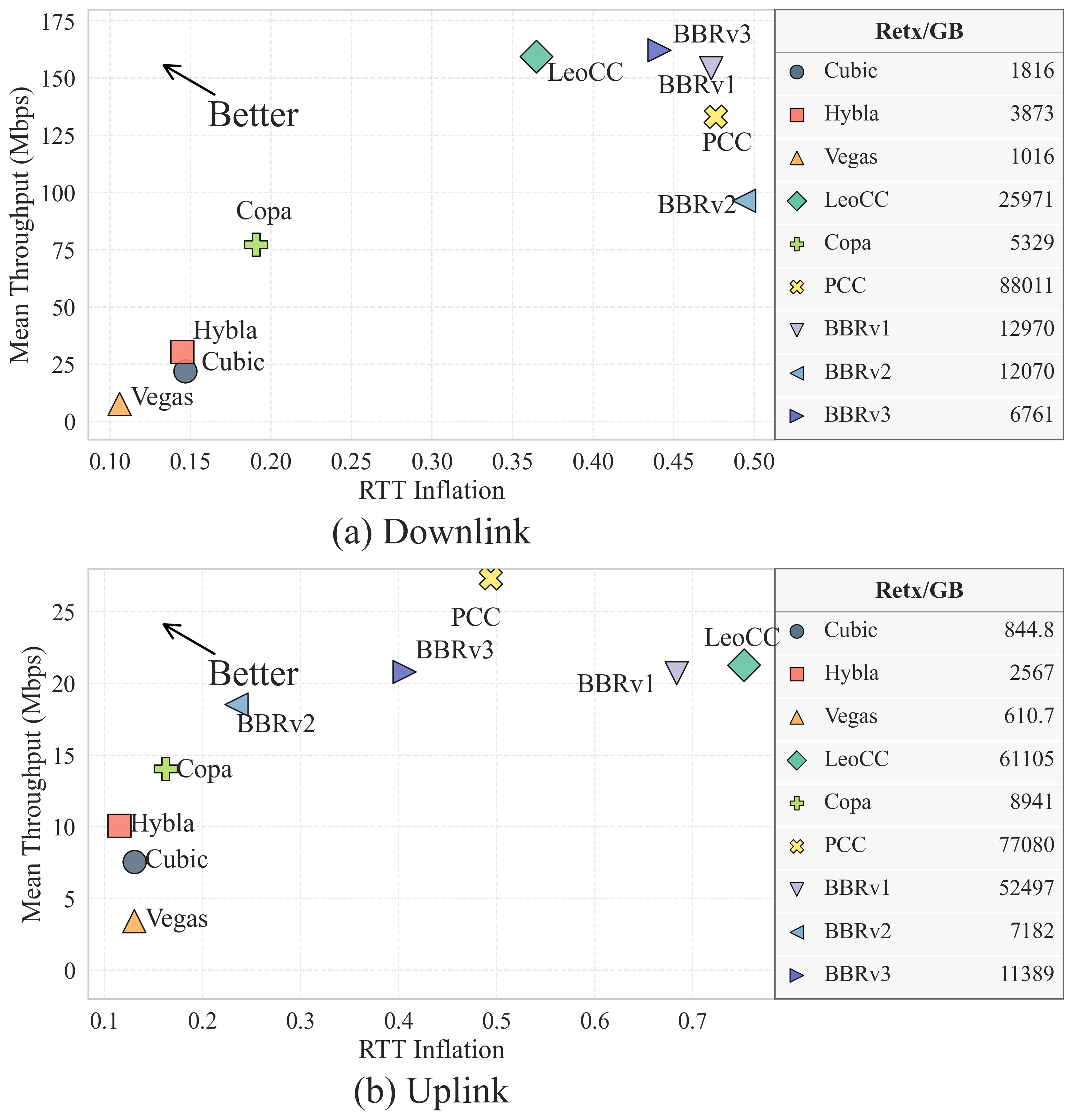}
  \caption{\textcolor{black}{Throughput vs RTT inflation of dedicated transmissions.}}
\label{fig:throughput_matrix_2}
\end{figure}

\begin{figure}[tb]
\centering
    \includegraphics[width=\linewidth]{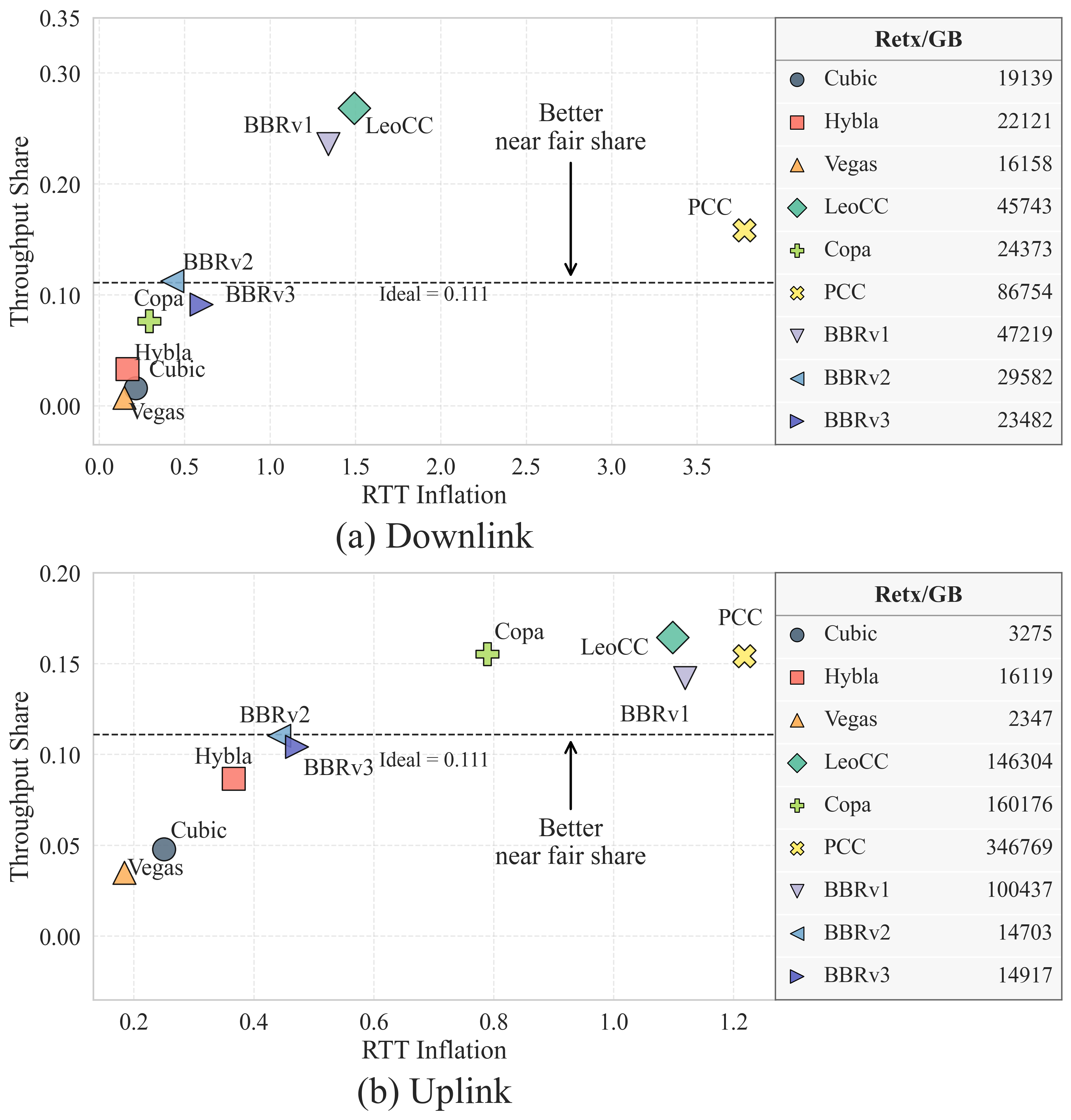}
  \caption{\textcolor{black}{Throughput share vs RTT inflation of concurrent transmissions.}}
\label{fig:throughput_share_matrix_2}
\end{figure}

% {\color{blue} Fig.~\ref{fig:throughput_matrix_2} shows that capacity-seeking schemes such as \gls{bbrv3}, LeoCC, PCC, \gls{bbrv1}, and \gls{bbrv2} achieve higher dedicated-flow throughput than conservative schemes such as Cubic, Hybla, Vegas, and Copa. However, PCC, LeoCC, and \gls{bbrv1} incur higher retransmission costs. In contrast, \gls{bbrv3} maintains high throughput with lower retransmissions, especially in downlink, and remains competitive in the more capacity-limited uplink. The terrestrial baseline in Appendix~\ref{appendixC} further shows that these behaviours are mainly caused by Starlink path dynamics.
% A throughput vs RTT inflation analysis accounting for the total 10 runs is presented in Fig.~\ref{fig:throughput_share_matrix_2}.  It highlights that under concurrent flows, Cubic, Hybla, and Vegas stay below fair share due to conservative delay/loss reactions. 
% LeoCC, PCC, and \gls{bbrv1} capture more bandwidth but increase \gls{rtt} inflation and retransmissions. \gls{bbrv2} and \gls{bbrv3} operate closer to fair share. Overall, \gls{bbrv3} does not simply maximize throughput; it provides a better fairness--delay--loss trade-off over Starlink.
Fig.~\ref{fig:throughput_matrix_2} summarizes the dedicated-flow behavior by jointly presenting mean throughput, \gls{rtt} inflation, and retransmissions per GB evaluation, encapsulating the data collected over the total runs. 
The results show that model-based or capacity-seeking schemes, including \gls{bbrv3}, LeoCC, PCC, \gls{bbrv1}, and \gls{bbrv2}, achieve substantially higher throughput than conservative loss- or delay-sensitive schemes such as Cubic, Hybla, Vegas, and Copa. 
However, this higher capacity capture often comes with increased retransmission cost, particularly for PCC, LeoCC, and \gls{bbrv1}. 
However, in the downlink, \gls{bbrv3} lies in the high-throughput region while maintaining lower retransmissions than the most aggressive alternatives, and its predecessors. 
In the uplink, where the Starlink path is more capacity constrained, \gls{bbrv3} remains competitive while avoiding the severe retransmission overhead observed for PCC, LeoCC, and \gls{bbrv1}. 
To get a better perspective of the Starlink constraints, we have presented a terrestrial network throughput comparison of the downlink and uplink paths in Appendix~\ref{appendixC},  Fig~\ref{fig.terres_performance}. 
The results further strengthen that \gls{bbrv3} provides a balanced dedicated-flow operating point over Starlink, sustaining high utilization without excessive loss penalties.

On the other hand, Fig.~\ref{fig:throughput_share_matrix_2} summarizes the concurrent-flow behavior by comparing throughput share, \gls{rtt} inflation, and retransmissions per GB against the ideal fair-share line. 
The results show that Starlink contention separates the \glspl{cca} into three operating regimes. 
Conservative schemes such as Cubic, Hybla, and Vegas remain below the fair-share line in both downlink and uplink because their loss or delay-sensitive responses limit their ability to compete under Starlink's dynamic network conditions. 
In contrast, LeoCC, PCC, and \gls{bbrv1} capture a larger share of the bottleneck capacity, especially in downlink, but this comes with substantially higher \gls{rtt} inflation and retransmission intensity, indicating aggressive bandwidth acquisition with increased loss exposure. 
\gls{bbrv2} and \gls{bbrv3} operate closer to the fair-share region and with moderate \gls{rtt} inflation in both downlink and uplink.
Importantly, \gls{bbrv3} maintains much lower retransmissions per GB than LeoCC, PCC, and \gls{bbrv1}, while avoiding the severe throughput starvation observed for conservative CCAs. 
Thus, under concurrent-flow operation, \gls{bbrv3}'s advantage is not maximum bandwidth capture, but a more balanced fairness, loss, and delay trade-off over the Starlink Internet.
\color{black}

\section{Modeling \gls{bbrv3} Starlink Internet}
\label{sec:fluid_model}

\subsection{\gls{bbrv3} Fluid Model} 
\label{subsec:fluid_model}

% For tractability~\cite{scherrer2022model}, we consider flows with large \glspl{bdp}, 
% \gls{bbr} implements bandwidth probing more frequently than in general cases, though it remains in the same time scale as Cubic. 
\gls{bbrv3} probing occurs every $62\cdot RTT_i^{min}$ for low \gls{rtt} flows, while for high \gls{rtt} flows, it happens at randomly selected intervals between 2 and 3 seconds. 
To eliminate stochasticity, let us define the time interval between sequential \texttt{ProbeBW} events for any flow $i \in \{1,\ldots,N\}$ in a parallel set of $N$ \gls{bbr} flows as~\cite{scherrer2022model}:
\begin{equation}
\bar{t}^{pbw} = \min\Big(62\cdot \text{RTT}^\text{{rtp}}, 2+ \frac{i}{N}\Big)
\end{equation}
where $\text{RTT}^\text{{rtp}}$ represents the estimated minimum propagation \gls{rtt}.

We introduce the indicator variable $\mathbb{I}_{dwn}$ as follows, which equals 1 when the \gls{bbr} flow attempts to reduce its inflight data:
\begin{equation}
    \mathbb{I}^{dwn} = 
    \begin{cases}
        1, & v(t) > \tfrac{5}{4}\bar{BDP} \;\lor\; p^{tot}(t) > 0.02, \\
        0, & \text{otherwise},
    \end{cases}
\end{equation}
where $v(t)$ represents the inflight volume, which evolves according to sender injection and network delivery rates, and $\bar{BDP}$ is the estimated \gls{bdp}. 
$p^{tot}$ is the end-to-end probability of packet drop due to link failure defined in Eq. (\ref{eq:total_prob}).
Similarly, $\mathbb{I}^{crs}$ indicates whether the flow is in cruising state:
\begin{equation}
    \mathbb{I}^{crs} = 
    \begin{cases}
        1, & v(t) \leq \bar{BDP}, \\
        0, & \text{otherwise},
    \end{cases}
\end{equation}
The pacing rate is modeled as:
\begin{equation}
    x^{pcg} = x^{btl}\Big(1 + \frac{\sigma(t_i^{pbw} - \text{RTT}^{\text{rtp}})(1-\mathbb{I}^{dwn}) - \mathbb{I}^{dwn}}{4}\Big)
    \label{eq:pacing_rate}
\end{equation}
where $x^{btl}$ represents the estimated bottleneck bandwidth. When $\mathbb{I}^{dwn} = 0$, the pacing rate increases to $\frac{5}{4}\cdot x^{btl}$; otherwise, it reduces to $\frac{3}{4}\cdot x^{btl}$.

Phase transitions are triggered by probing observations. $\mathbb{I}^{\text{dwn}}$ activates when $v(t) > \tfrac{5}{4} \cdot \bar{\text{BDP}} \;\lor\; p^{tot}(t) > 0.02$, and deactivates once inflight data drops to the conservative target $\hat{\text{BDP}} = \min\left(\bar{\text{BDP}},\, 0.85\,\text{BDP}^{\text{hi}}\right)$, where $\text{BDP}^{\text{hi}}$ represents the \texttt{inflight\_hi} long-term upper bound. 
This ensures queue clearance without pipe under utilization. 
% An interplay of the inflight limit variables using Sydney downlink data is illustrated in Fig. \ref{fig.fluid_model}. 
The transition dynamics are modeled as:
\begin{align}
\Delta^{\text{dwn}} =\; & (1 - \mathbb{I}^{\text{crs}})(1 - \mathbb{I}^{\text{dwn}}) \, \sigma(t^{\text{pbw}} - \text{RTT}^{\text{rtp}}) \nonumber \\
& \times \min\left(\sigma\left(v - \frac{5 \bar{\text{BDP}}}{4}\right) + \sigma(p^{tot} - 0.02),\, 1\right) \nonumber \\
& - \mathbb{I}^{\text{dwn}} \, \sigma(\hat{\text{BDP}} - v)
\end{align}
Deactivation of $\mathbb{I}^{\text{dwn}}$ directly activates $\mathbb{I}^{\text{crs}}$, which is disabled in \texttt{ProbeBW} states:
\begin{equation}
\Delta^{crs} = -\Delta^{dwn} - \sigma (t^{pbw} - \bar{t}_i^{pbw})\mathbb{I}_i^{crs}
\end{equation}
The \texttt{ProbeBW} congestion window is modeled as \cite{bbrv3, scherrer2022model}:
\begin{equation}
    w^{pbw} = 
    \begin{cases}
        \min \Big( 2\bar{\text{BDP}},\; \text{BDP}^{\text{lo}}\Big), 
        & \mathbb{I}^{crs} = 1 \;\; (\texttt{inflight\_lo}), \\
        \min \Big( 2.5\bar{\text{BDP}},\; \text{BDP}^{\text{hi}} \Big), 
        & \mathbb{I}^{crs} = 0 \;\; (\texttt{inflight\_hi}),
    \end{cases}
\end{equation}

while in \texttt{ProbeRTT} state, it is:
\begin{equation}
  w^{prt} = \frac{\bar{\text{BDP}}}{2}
\end{equation}

\color{black}

\begin{figure}[t]
\centering
    \includegraphics[width=\columnwidth]{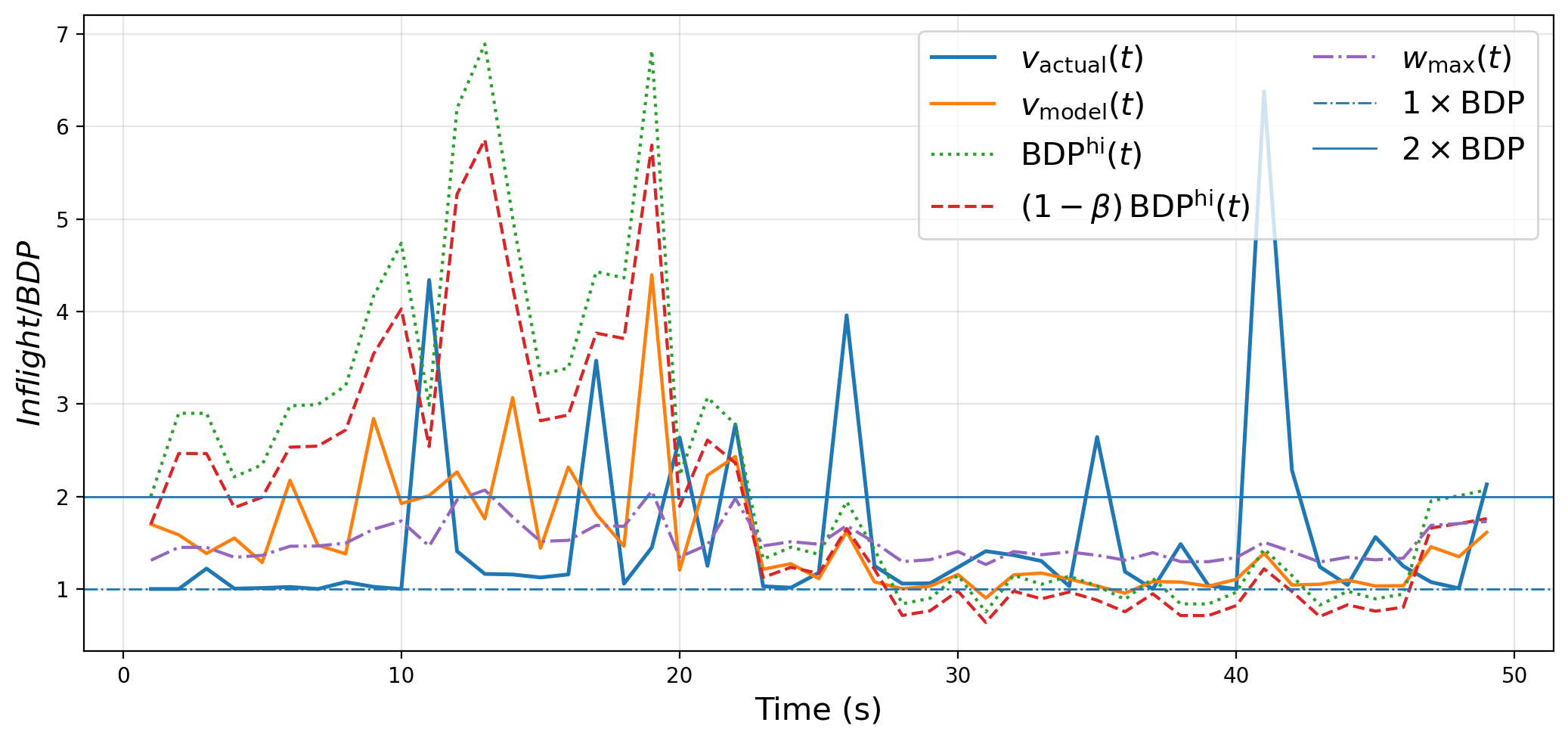}
  \caption{\textcolor{black}{Visualization of BBR fluid model inflight limits}}
\label{fig.fluid_model}
\end{figure}

The inflight dynamics can be defined as:
\begin{equation}
\frac{dv(t)}{dt} = x^{pcg}(t) - x(t),
\end{equation}
where $x^{pcg}(t)$ (Eq. \ref{eq:pacing_rate}) and $x(t)$ is the network delivery rate.
Under quasi-steady conditions, this yields:
$v_{\text{actual}}(t) \approx x(t)\text{RTT}(t)$,
encapsulating a measurement-based estimate of inflight, where $x(t)$ is the measured delivery rate.
To obtain a tractable model, we approximate the dynamics using a first-order relaxation:
\begin{equation}
\frac{dv_{\text{model}}(t)}{dt}
=
\lambda \left( v_{\text{target}}(t) - v_{\text{model}}(t) \right),
\label{eq:first_order_relaxation}
\end{equation}
where $v_{\text{target}}(t)$ is determined by the ProbeBW state and the inflight bounds derived above.
Figure~\ref{fig.fluid_model} provides a validation of the proposed BBRv3 fluid model using a Sydney to local terminal downlink data capture. 
The long-term inflight bound $\text{BDP}^{hi}(t)$ is derived from loss events inferred from retransmissions, enabling direct comparison between the measured inflight $v_{\text{actual}}(t)$ and the model-predicted inflight $v_{\text{model}}(t)$ under network-driven constraints.

In the probing phase ($t \approx 0$--20 s), the measured inflight $v_{\text{actual}}(t)$ exhibits large excursions above $1\times\bar{\text{BDP}}$. 
Predicted inflight $v_{\text{model}}(t)$ follows the upper envelope defined by $\text{BDP}^{hi}(t)$, capturing the dominant probing behavior of BBRv3.
Around $t \approx 20$ s, both $v_{\text{actual}}(t)$ and $\text{BDP}^{hi}(t)$ show a sharp decrease due to a change in network conditions. 
The model reproduces this transition through loss-driven updates of $\text{BDP}^{hi}(t)$, demonstrating its ability to capture time-varying path effects.
For $t \gtrsim 25$ s, both measured and modeled inflight converge toward the headroom-controlled region $(1-\beta)\text{BDP}^{hi}(t); (\beta = 0.15  = \text{queue draining margin} / \text{safety headroom})$ \cite{bbrv3} and remain close to $1\times\text{BDP}$, indicating consistent steady-state behavior.
We conjecture that the short-term spikes in $v_{\text{actual}}(t)$ arise from transient effects such as RTT variability and ACK compression, which are not explicitly modeled.

\vspace{-4mm}

\subsection{Measurement-Based Interpretation of \gls{bbrv3} Dynamics}
\label{subsec:results_with_model}

\begin{figure}[htb]
\centering

\subfloat[Downlink \gls{bbrv3} transmission stream\label{fig:downlink_measurement_validation}]{
    \includegraphics[width=\columnwidth]{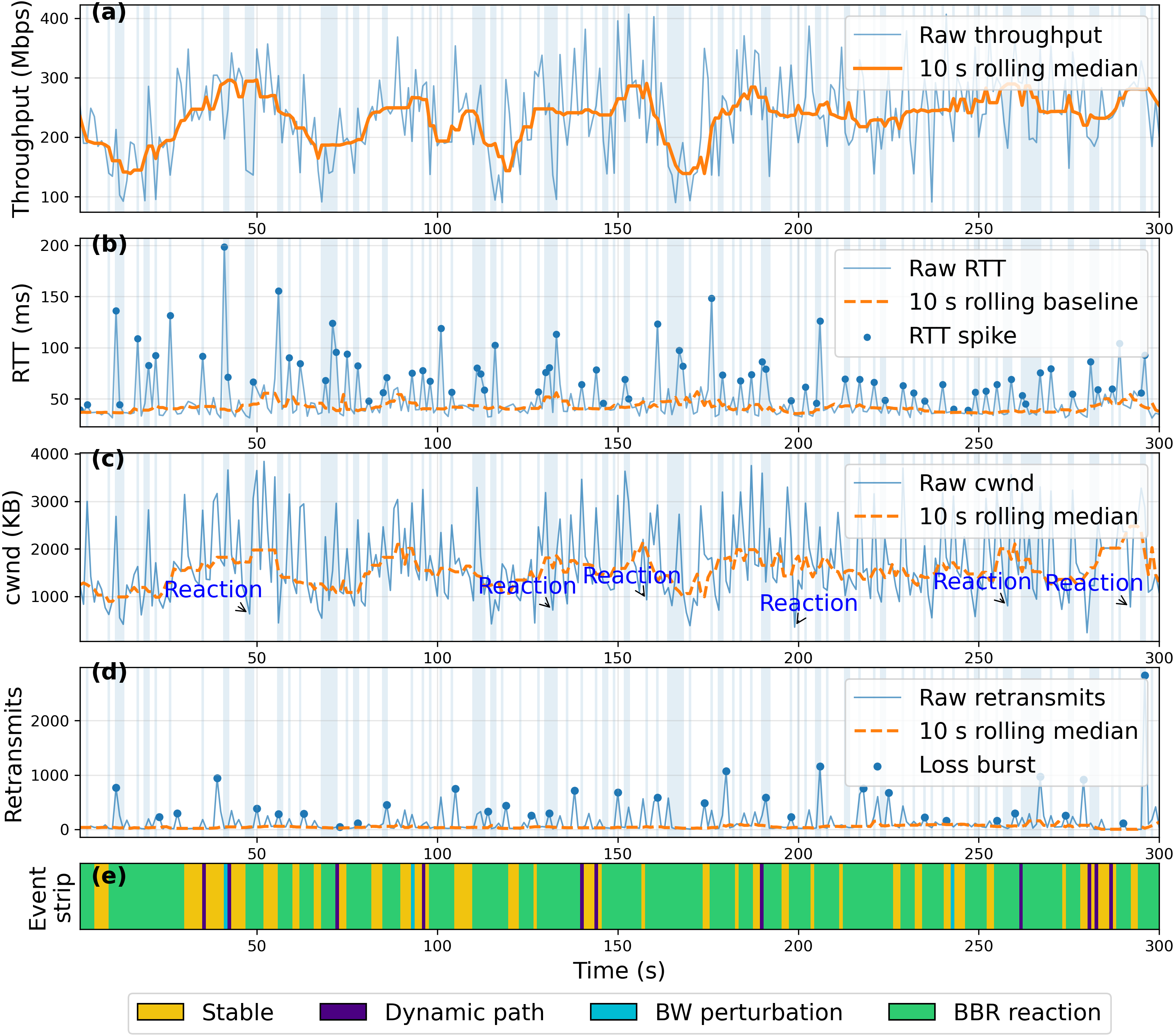}
}

\vspace{2mm}

\subfloat[Uplink \gls{bbrv3} transmission stream\label{fig:uplink_measurement_validation}]{
    \includegraphics[width=\columnwidth]{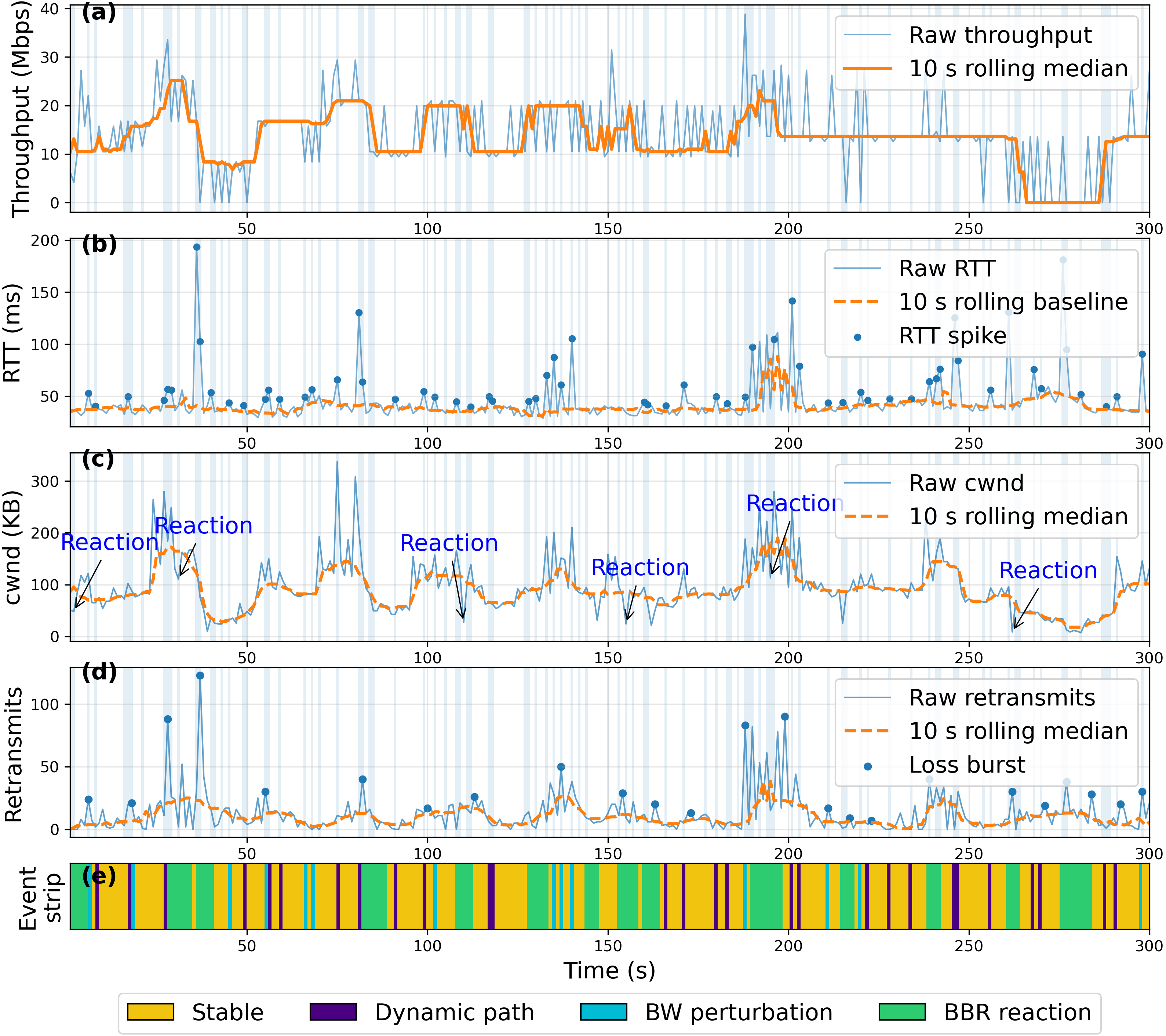}
}

\caption{\textcolor{black}{Measurement validation through dedicated downlink and uplink \gls{bbrv3} transmission streams.}}
\label{fig:measurement_validation}
\end{figure}

\gls{bbr} continuously estimates maximum delivery rate ($B_{\theta}$) and minimum RTT ($\text{RTT}_{\min}$), setting its sending rate as $R_{\text{sending}} = G \cdot B_{\theta}$, where $G$ is a phase-dependent gain factor. 
We present a validation of the analysis using Sydney captures, encapsulating these properties. 
A \gls{bbr} reaction window is identified if an \gls{rtt} spike or bandwidth drop is followed by a congestion window drop within the reaction window.
Measurement-based validation of the proposed Starlink–\gls{bbrv3} coupled model using Sydney captures is illustrated in Fig.~\ref{fig:measurement_validation} for both downlink and uplink.

The \gls{rtt} variation due to satellite motion, inter-satellite routing, and queue accumulation, and can be expressed as:
\begin{equation}
    RTT(t)=\frac{2d_s(t)}{c}+\sum_i \frac{d^I_i(t)}{c}
+\frac{q(t)}{\mathbb{C}(t)}
\label{eq:rtt}
\end{equation}
where $d_s(t)$ is the satellite distance and $d^I_i$ is the signal traveling distance of $i^{\text{th}}$ \gls{isl}, $q(t)$ is the instantaneous queue occupancy, $C(t)$ is the instantaneous bottleneck capacity, and $c$ is the speed of light. 
Therefore, the observed RTT excursions indicate temporary changes in propagation distance or queuing delay.
Since the sender generates a continuous load, the measured throughput reflects the instantaneous effective capacity: 
\begin{equation}
    C_{\mathrm{eff}}(t)=\min[C(t),C_{\mathrm{ISL}}(t)](1-p^{tot}(t))
\end{equation}
where $C_{\mathrm{ISL}}$ is the \gls{isl} capacity and $p^{tot}$~(Eq.\eqref{eq:total_prob}) is the path uncertainty probability (See Appendix~\ref{appendix:link_analysis} for further information on Starlink capacity modeling). 
% recall Eq. \eqref{eq:capacity} for instantaneous capacity $C(t)$, Eq. \eqref{eq:isl_capacity_final} for \gls{isl} capacity, and Eq. \eqref{eq:total_prob} for path uncertainty probability $p^{tot}(t)$. 
Therefore, the retransmission bursts shown in Fig.~\ref{fig:measurement_validation} are consistent with increases in $p^{tot}(t)$ or queue overflow.
Intervals, where RTT excursions coincide with throughput degradation, are classified as bandwidth perturbation events, which are highlighted in Fig.~\ref{fig:measurement_validation} respective subplots.

Under the proposed fluid model, the behavior observed in Fig.~\ref{fig:measurement_validation} can be interpreted through the state variables $\bar t^{pbw}$, $\mathbb I^{dwn}$, and $\mathbb I^{crs}$.
The RTT excursions and retransmission bursts indicate intervals in which the Starlink path departs from its nominal operating point, corresponding to conditions where either the inflight
process exceeds $\tfrac{5}{4}\bar{BDP}$ or the end-to-end packet-drop
probability becomes significant.
These conditions correspond exactly to the scenarios in which $\mathbb I^{dwn}$ is activated within the fluid model.
Once $\mathbb I^{dwn}=1$, the pacing rule (Eq.\eqref{eq:pacing_rate}) reduces
$x^{pcg}$ from its probing level toward the conservative level, which
appears as throughput degradation during disturbed intervals.

The congestion-window variations in Fig.~\ref{fig:measurement_validation} provide the measurement-side manifestation of the inflight limits $w^{pbw}$ and $w^{prt}$.
When the path remains stable and $\mathbb I^{crs}=1$, the sender operates
near the cruising/probing envelope, when RTT inflation and retransmission
bursts occur, the transition dynamics $\Delta^{dwn}$ and $\Delta^{crs}$
drive the flow away from cruising and toward a reduced inflight state.
Thus, producing the repeated congestion window contractions marked as reaction intervals.
Accordingly, the event strips in Fig.~\ref{fig:measurement_validation} summarize the phase evolution predicted by the fluid model, dynamic-path disturbance
$\rightarrow$ bandwidth perturbation $\rightarrow$ down transition
$\rightarrow$ \gls{bbrv3} reaction.
Moreover, the disturbance spacing is on the order of tens of seconds,
which is comparable to $\bar t^{pbw}$, confirming that \gls{bbrv3} over Starlink
must be analyzed as a coupled non-stationary system rather than as a
conventional fixed-bottleneck path.

\color{black}

\vspace{-2mm}

\subsection{Queuing Analysis}
\label{subsec:queuing}

The end-to-end latency comprises baseline propagation delay and variable queuing delay as expressed in Eq. (\ref{eq:rtt}). 
% Assuming negligible queuing in the reverse path for TCP ACKs, we focus on RTT as measured by the TCP sender, with $q \approx \max(0, \text{RTT} - \text{RTT}_{\text{base}})$. We estimate $\text{RTT}_{\text{base}}$ using a rolling $5^{\text{th}}$ percentile filter over a 120-second window: $q = \text{Quantile}_{0.05}\!\left( \text{RTT}_{i-w+1}, \ldots, \text{RTT}_i \right)$, where $w$ represents the sample window length. This approach minimizes congestion and jitter effects, tracks the RTT floor corresponding to empty or near-empty queues, and adapts gradually to path changes such as satellite handovers or orbital movements.
To characterize queuing behavior over Starlink downlinks, we apply an $M/G/1$ queue approximation to our empirical data. With MSS $m=1500$ bytes and \textcolor{black}{sampling window ($w$) of 15~s}, we derive service rate $\mu = \frac{b}{8\cdot m}$ [packets/s] from observed throughput $b$. Since packet arrivals are not directly observable, we approximate the arrival rate $\lambda$ using the observed service rate and the queue occupancy fraction. Defining $X_j = \mathbb{I}(q_j > 0)$ as a queue occupancy indicator, we estimate the arrival rate as $\lambda_i = \mu_i \cdot \bar{X}_i$, where:
\begin{equation}
    \bar{X}_i = \frac{1}{w} \sum_{j=i-w+1}^i X_j
\end{equation}
This assumes arrivals occur at the service rate during busy periods and proportionally less during idle periods, consistent with the utilization relation $\rho = \lambda/\mu$.

In an $M/G/1$ queue, the mean waiting time follows the Pollaczek–Khinchine formula:
\begin{equation}
W_q = \frac{\lambda \, \mathbb{E}[S^2]}{2\left(1 - \rho\right)}
\end{equation}
where $\rho = \lambda \, \mathbb{E}[S]$ represents utilization and $S$ is the service time distribution. With $\mathbb{E}[S] = 1/\mu$ and $\mathbb{E}[S^2] = (1+c_s^2)(\mathbb{E}[S])^2$, where $c_s^2$ is the squared coefficient of variation of service time (assumed $c_s^2 = 1$ for exponential service times), the average queue size follows Little's Law as $Q = \lambda \, W_q$.

\begin{figure}[t]
\centering
    \includegraphics[width=\columnwidth]{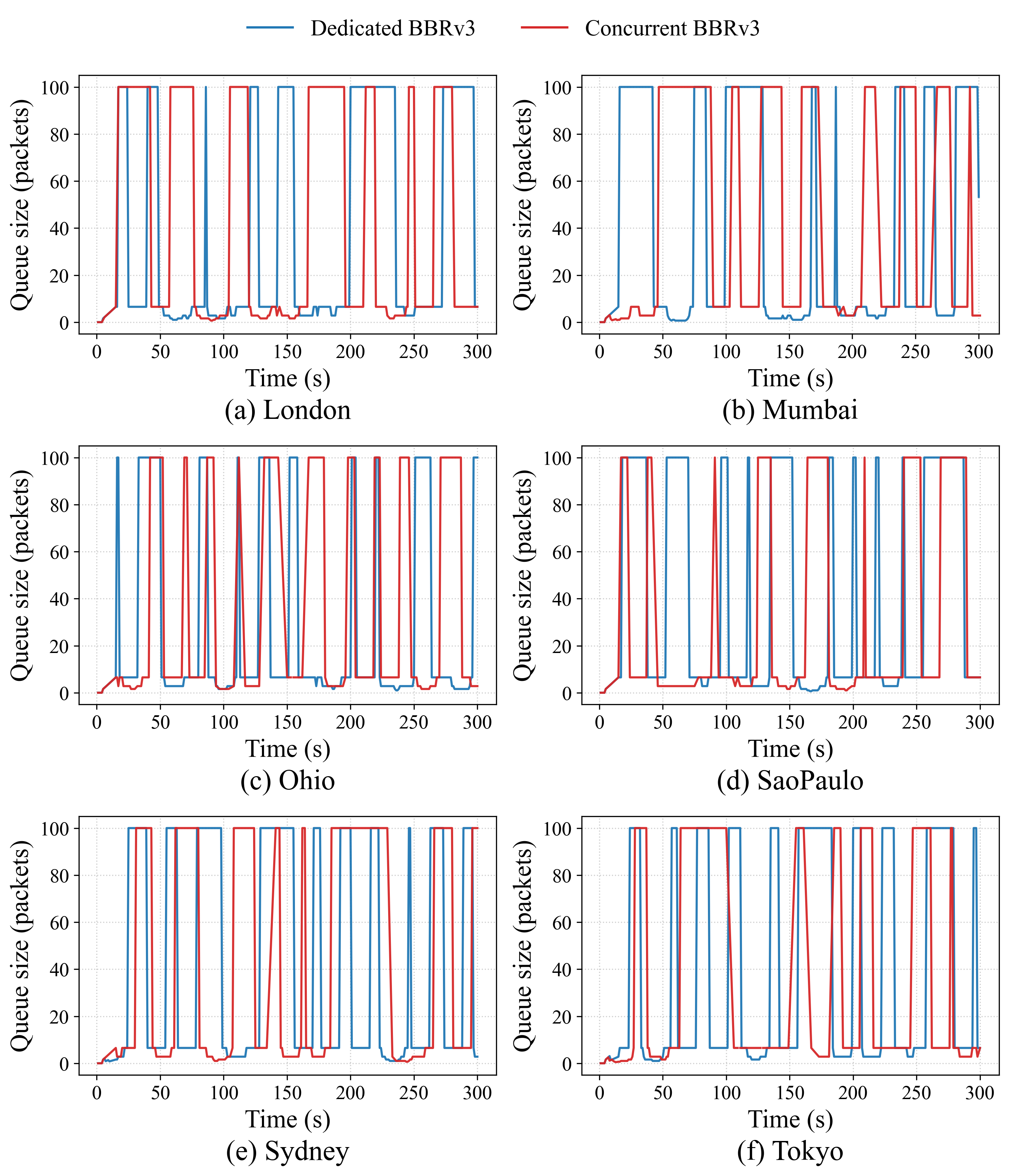}
  \caption{\textcolor{black}{Queue buildup in dedicated and concurrent downlink BBR streams for different server locations over the Starlink network}}
\label{fig:downlink_queue_models}
\end{figure}

\color{black}
Fig.~\ref{fig:downlink_queue_models} depicts the temporal evolution of the estimated queue size under \gls{bbrv3} for downlink flows across the six geographically distributed Starlink paths. 
A prominent feature across all locations is the recurrent saturation of the queue at approximately $Q \approx 100$ packets. 
This characteristic emerges implicitly from the $M/G/1$ formulation, specifically through the bounded utilization constraint $\rho_i = \lambda_i \mathbb{E}[S_i] \leq \rho_{\max}$. 
As $\rho_i \to 1$, the Pollaczek--Khinchine relation
causes the expected waiting time to grow rapidly, and consequently, the queue size
$Q_i = \lambda_i W_{q,i}$
approaches a finite but large value determined by the imposed upper bound on $\rho_i$ ($\rho_{\max} = 0.995$), yielding the observed saturation near 100 packets.

Both dedicated and concurrent \gls{bbrv3} flows show frequent transitions between $\rho \rightarrow 1$ and average queue buildup of around 7 packets, as depicted in Fig.~\ref{fig:downlink_queue_models}. 
Thus confirming that \gls{bbrv3} frequently drives the Starlink downlink close to full utilization.
However, the concurrent traces tend to show more irregular timing and shorter low-queue intervals in several cities. 
This is expected because, under concurrent operation, \gls{bbrv3} shares the bottleneck with other active congestion-control flows, so the aggregate offered load keeps the bottleneck closer to saturation even when the \gls{bbrv3} flow temporarily reduces its sending rate. 
In contrast, the dedicated \gls{bbrv3} traces show clearer drain periods in some cities, because the queue evolution is governed mainly by a single \gls{bbrv3} control loop. 
Hence, the result indicates that \gls{bbrv3}’s downlink behavior over Starlink is dominated by repeated transitions between queue-draining and near-saturation phases.
At the same time, concurrent traffic increases the persistence and irregularity of the saturated queue state.
\color{black}

\vspace{-5mm}

\color{black}
\subsection{BBR-v3 Fairness Analysis}
\label{subsec:fairness}

\begin{figure}[tb]
\centering
    \includegraphics[width=\columnwidth]{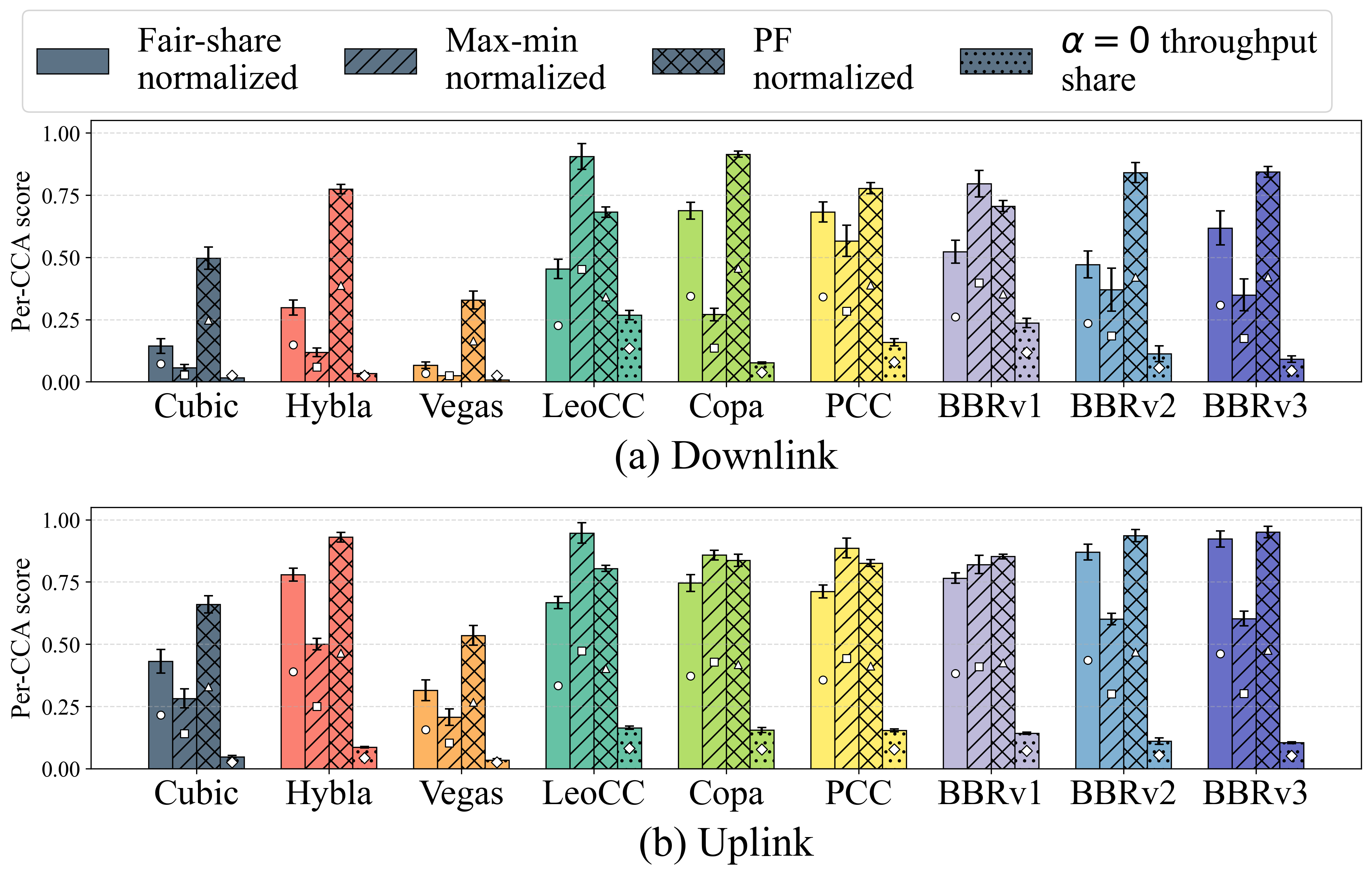}
  \caption{\textcolor{black}{Normalized TCP fairness indexes of the evaluated CCAs over the Starlink Internet. }}
\label{fig:per_cca_standard_fairness}
\end{figure}

Considering the uplink and downlink data collected under concurrent-flow scenario from the distributed Starlink testbed, we evaluate \gls{bbrv3} fairness relative to the other \glspl{cca}. 
Normalized Jain fairness index is defined as:
\begin{equation}
J=\frac{\left(\sum_i T_i\right)^2}{N\sum_i T_i^2},
\end{equation}
where $T_i$ denote the mean throughput of \gls{cca} $i$, $N$ the number of competing \glspl{cca}. The normalized max--min closeness is:
\begin{equation}
F_i^{\mathrm{MM}}=\frac{T_i}{\max_j T_j},
\end{equation}
and normalized Proportional-Fairness~(PF):
\begin{equation}
F_i^{\mathrm{PF}}=\min\left(\frac{C_i}{S^\star},\frac{S^\star}{C_i}\right),
\qquad
C_i=\frac{\log(1+T_i)}{\sum_j \log(1+T_j)},
\end{equation}
where $S_i=T_i/\sum_{j=1}^{N}T_j$ its achieved throughput share, $S^\star=1/N$. 
The $\alpha=0$ throughput-maximization share is: $F_i^{\alpha=0}=S_i$, were calculated for each \gls{cca} under this evaluation.  
These metrics allow \gls{bbrv3} to be assessed not only by its throughput share, but also by how closely it approaches equal sharing, how far it is from the dominant flow, and how its proportional-fairness contribution compares with the ideal allocation.

\gls{bbrv3} fairness evaluation results under above metrics are illustrated in Fig.~\ref{fig:per_cca_standard_fairness}. 
In the downlink case, \gls{bbrv3} remains competitive and achieves one of the highest PF-normalized scores, indicating that its proportional-fairness contribution remains close to the ideal allocation. 
However, its fair-share score, max--min score, and $\alpha=0$ throughput share are lower than the more aggressive high-throughput \glspl{cca}, particularly LeoCC, PCC, \gls{bbrv1}, and \gls{bbrv2}. 
This shows that \gls{bbrv3} does not consistently dominate the downlink bottleneck, instead, it operates in a less aggressive regime while still maintaining strong proportional-fairness behavior. 
In the uplink case, \gls{bbrv3} exhibits a stronger overall fairness profile, achieving a high fair share and PF-normalized scores while maintaining a throughput share close to the ideal allocation. 
Although its max--min score remains below that of LeoCC, PCC, Copa, and \gls{bbrv1}. 
These combined metrics suggest that \gls{bbrv3} coexists effectively with concurrent \glspl{cca} flows without persistently over-capturing the shared Starlink path. 
Thus, highlighting its favorable fairness--efficiency tradeoff and robust coexistence characteristics across both Starlink directions.

\begin{figure}[tb]
\centering
    \includegraphics[width=\columnwidth]{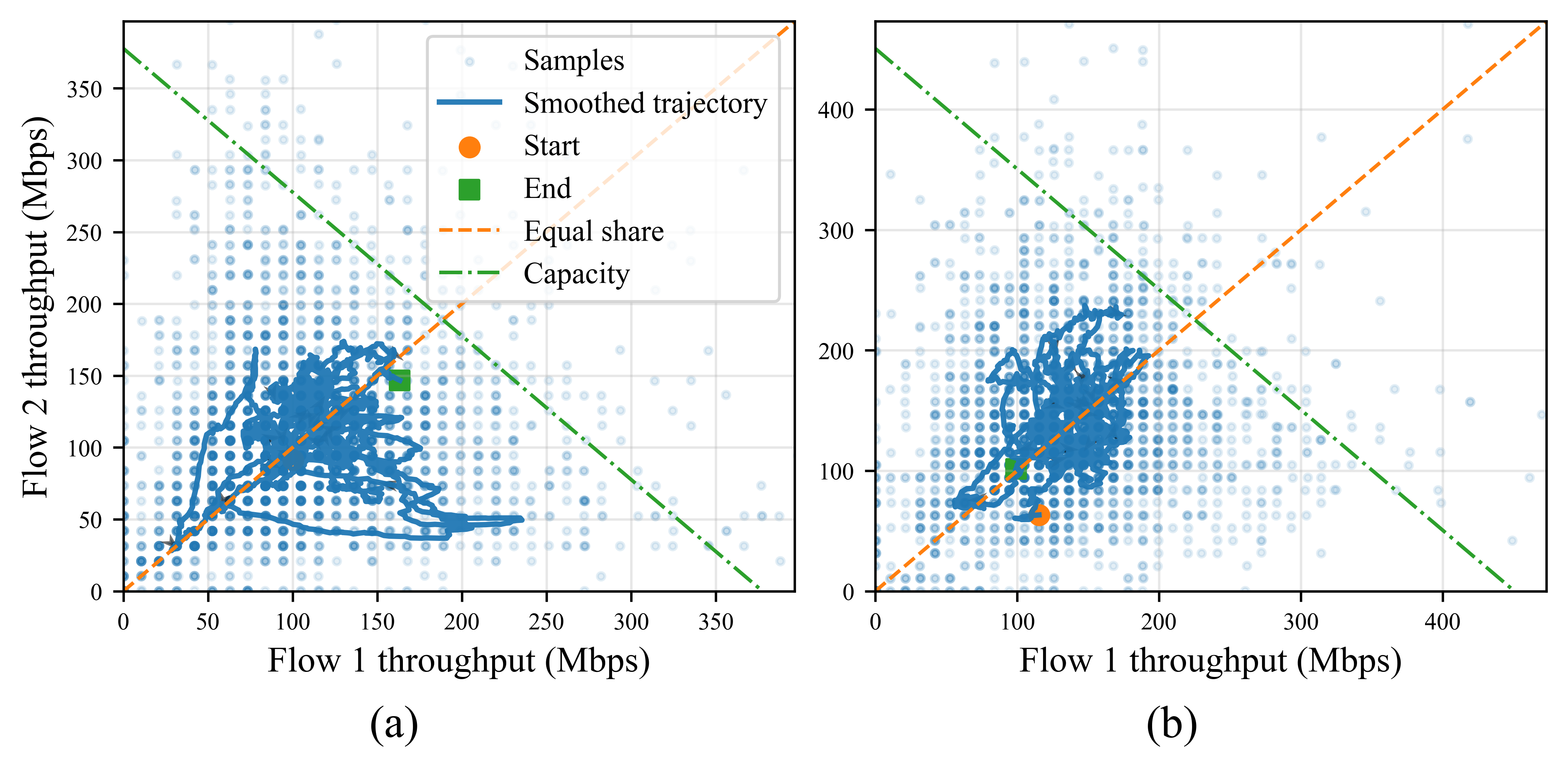}
    \vspace{-5mm}
  \caption{\textcolor{black}{Throughput realized by \gls{bbrv3} connections (a) Simultaneous flows (b) Staggered flows. }}
\label{fig:bbr3_fairness}
\end{figure}

To further evaluate the fairness characteristics of \gls{bbrv3} over Starlink, we conducted dual-flow experiments between an Amazon EC2 instance in Sydney and the local Starlink terminal over the downlink path. 
Two \gls{bbrv3} \gls{tcp} flows were generated under simultaneous and staggered arrival conditions, with the latter introducing a 30~s offset between flows. 
Their throughput dynamics were analyzed using the phase-plane representation which is presented in Fig.~\ref{fig:bbr3_fairness}.
In the illustration, each point denotes the instantaneous throughput pair of the two flows. 
Furthermore, fairness was quantified using the above detailed four complementary allocation metrics: Jain's fairness index, max--min fairness ratio, PF score, and the $\alpha=0$ throughput-maximization utility for both \gls{bbrv3} flows. 
In the simultaneous case, \gls{bbrv3} achieves balanced sharing, with a mean instantaneous Jain's index of 0.949, PF score of 0.946, and a mean flow throughput ratio of 0.992, indicating that the two flows obtain nearly symmetric rates. 
The max--min score is 0.761, showing that short-term throughput imbalance still occurs even though the long-term mean rates are close. 
In the staggered case, the mean instantaneous Jain's index remains relatively high at 0.909 and the PF score is 0.907, but the max--min score decreases to 0.637 and the throughput ratio increases to 1.258. 
This indicates that the earlier flow retains a measurable advantage after the second flow joins. 
However, the $\alpha=0$ throughput-maximization utility increases from 197.30~Mbps in the simultaneous case to 242.52~Mbps in the staggered case, suggesting that the staggered scenario improves aggregate throughput at the cost of reduced fairness. 
Overall, these results show that \gls{bbrv3} maintains stable two-flow coexistence over the Starlink downlink path, but its fairness-efficiency tradeoff is sensitive to flow arrival order.

\color{black}

\vspace{-2mm}

\subsection{New Insights on \gls{bbr} Performance}

Our experimental results demonstrate that \gls{bbrv3} achieves high throughput in dedicated flows while maintaining a more balanced fairness, retransmission, and delay trade-off than aggressive alternatives such as PCC, LeoCC, and \gls{bbrv1} under concurrent-flow operation. This stems from its model-driven approach of explicitly estimating bottleneck bandwidth and round-trip propagation delay, rather than reacting to loss events. While Starlink's dynamic conditions limit loss-based algorithms such as Cubic and Hybla and remain below the fair-share line in concurrent flows, \gls{bbrv3} sustains near-capacity utilization without incurring the excessive retransmission overhead observed for more aggressive \glspl{cca}. This balance between high utilization, moderate \gls{rtt} inflation, and low loss exposure distinguishes \gls{bbrv3} as a well-suited \gls{cca} for both dedicated and concurrent Starlink traffic.

\color{black}

\gls{bbrv3}'s exponential startup mechanism achieves rapid bottleneck utilization, contrasting sharply with the prolonged convergence characteristics of additive increase algorithms. Its periodic bandwidth probing through \texttt{ProbeBW\_UP} with pacing gain exceeding unity ensures swift discovery of available capacity, while competing \glspl{cca} remain constrained by conservative additive increase mechanisms. By intelligently bounding inflight data to the estimated \gls{bdp} via \texttt{inflight\_hi} and \texttt{inflight\_lo} parameters, \gls{bbr} effectively mitigates persistent queue buildup and bufferbloat.
Thus, maintaining comparable \gls{rtt} variance to benchmark \glspl{cca} despite higher throughput. Furthermore, \gls{bbrv3}'s tolerance for approximately 2\% loss per \gls{rtt} during probing prevents the catastrophic rate collapse characteristic of traditional \glspl{cca}.

\color{black}
The higher retransmission counts observed with \gls{bbrv3} stem from its model-based probing strategy, which periodically paces above the estimated bottleneck bandwidth to refine its bandwidth model. 
Unlike loss-based algorithms that treat packet loss as a catastrophic signal requiring multiplicative window reduction, \gls{bbrv3} uses it as an informational cue, allowing transmission to continue near capacity. 
\color{black}
Additionally, \gls{bbr}'s ACK aggregation modeling can generate transient bursts that exceed buffer capacity, causing clustered retransmissions. 
However, these behaviors also indicate that the default \gls{bbrv3} configuration is not fully optimized for Starlink's non-stationary capacity, \gls{rtt} shifts, and handover dynamics. 
As shown in Appendix~\ref{appendixD}, fixed pacing and congestion-window gains introduce an inherent utilization--retransmission trade-off, suggesting that a Starlink-oriented \gls{bbrv3} configuration could better balance throughput, fairness, and loss exposure.

\color{black}

% \begin{figure}[h]
% \centering
%     \includegraphics[width=\linewidth]{Images/starlink_sequential_cca_all_higher_better_score_heatmap_no_rwnd.png}
%   \caption{Heat Map}
% \label{fig:dedicated_heatmap}
% \end{figure}

%%%%%%%%%%%%%%%%%%%%%%%%%%%%%%%%%%%%%%%%%%%%%%%%%%%%%%%%%%%%%%%%%%%%%%%%%%%%%%%%%%%%%%%

\vspace{-2mm}

\section{Conclusions}
\label{sec:Conclusions}
In this paper, we present a globally distributed experimental evaluation of Google's \gls{bbrv3} over SpaceX's Starlink \gls{leo} satellite network. 
\color{black}
Using our six-city testbed, we evaluate \gls{bbrv3} performance against eight \glspl{cca}: Cubic, Hybla, Vegas, LeoCC, Copa, PCC, \gls{bbrv1}, and \gls{bbrv2}. 
The results demonstrate that \gls{bbrv3}'s advantage is not aggressive bandwidth capture, but a more balanced fairness, retransmission, and delay trade-off over the Starlink Internet.
Our mathematical modeling of Starlink's network, incorporating atmospheric effects and \gls{leo} dynamics, provides a robust framework for understanding transport layer behavior over the Starlink Internet. 
Furthermore, we introduce a fluid model, present $M/G/1$ queue investigation and fairness analysis of \gls{bbrv3} over the Starlink network, expanding the understanding of how it operates over the world's largest \gls{leo} constellation.\color{black}
As \gls{leo} satellite constellations continue rapid deployment, \gls{bbrv3} represents a compelling solution for maximizing performance in high-latency, variable-bandwidth environments. 
Future work will focus on adapting \gls{bbr} parameters specifically for satellite characteristics and developing new approaches that balance throughput maximization with retransmission efficiency.

% \section*{Acknowledgement}
% This work is supported by SmartSat CRC, whose activities are funded by the Australian Government’s CRC Program.

%%%%%%%%%%%%%%%%%%%%% End Section %%%%%%%%%%%%%%%%%%%%%%%%%%%%%%%%%
\vspace{-2mm}
\bibliographystyle{IEEEtran} %BST
\bibliography{mybib}

\appendices

\section{Link Failure and Packet Drop Probability in Starlink Transmission}
\label{appendix:link_analysis}

\vspace{-3mm}
\subsection{Starlink Channel Model}
\label{subsec:satellite_to_ground_link}

We model the Starlink communication channel as follows. The distance between a \gls{leo} satellite and ground terminal at time $t$ is given by:
\begin{equation}
d_s(t) = \sqrt{R^2+a^2-2Ra\sin \Big(\theta(t) + a\sin\Big[\frac{R}{a}\cos\theta(t)\Big] \Big)}
\label{eq:dsg}
\end{equation}
where $\theta(t)$ represents the elevation angle, $R$ is Earth's radius, and $a=R+h$ with $h$ denoting the \gls{leo} altitude \cite{al2023deep}. The corresponding free-space propagation loss is:
\begin{equation}
L^{s}_{fs}(t) = {\left(\frac{4\pi d_s(t)}{\lambda_s}\right)}^{2}
\end{equation}
where $\lambda_s$ is the signal wavelength \cite{khan2022rate}.
Signal attenuation encompasses multiple environmental factors. Cloud attenuation follows $L_\mathrm{c}^{s} = 10^{K_{1}M^{l}d_{c}/10}$, where $K_{1}$ is the attenuation coefficient, $M^{l}$ is liquid water density, and $d_{c}$~(km) is the signal traveling distance through clouds. 
Rain attenuation is modeled as $L_\mathrm{r}^{s} = 10^{{a\bar{R}^{b}d_{r}}/10}$, with $\bar{R}$ representing rain rate, $d_{r}$~(km) the path through rain, and parameters $a,b$ determined by frequency and polarization \cite{ITU_AtmosphericLoss2019}. 
Scintillation and multipath fading contribution is defined under \gls{itu} guidelines as $L_\mathrm{s}^{s} = 10^{a'\cdot \sigma'}$, where $a'$ and $\sigma'$ are defined in \cite{ITU‑R‑P.618‑14}, while gaseous attenuation $L_\mathrm{g}^{s}$ follows the \gls{itu} definition in~\cite{ITU_R_P.676_13}.
The Doppler shift due to relative motion is $f_d = {\frac{v}{c}}{f_c}$, where $v$ is relative velocity, $c$ is light speed, and $f_c$ is carrier frequency. The receiver antenna gain is $G_\mathrm{r}^{s} ={4 \pi a'_r}/{\lambda_d^{2}}$, where $a'_r$ is receiver aperture area and $\lambda_d = {c}/({f_c \pm f_d})$ \cite{weththasinghe2024optimising}. 
With transmitter gain $G_\mathrm{t}^{s}$ and respective antenna efficiencies $\eta_{t}^{s},\eta_{r}^{s}$, the received power at time $t$ is:
\begin{equation}
P_\mathrm{r}^{s}(t) =
\frac{P_\mathrm{t}^{s} \cdot \eta_{t}^{s} \cdot \eta_{r}^{s} \cdot G_\mathrm{t}^{s} \cdot G_\mathrm{r}^{s}}
{L_\mathrm{fs}^{s}(t) \cdot L_\mathrm{c}^{s} \cdot L_\mathrm{r}^{s} \cdot L_\mathrm{s}^{s} \cdot L_\mathrm{g}^{s}}
\label{link_budget_satellite}
\end{equation}
yielding an \gls{snr} of $SNR^{S}(t) = {P_\mathrm{r}^{s}(t)}/{N_{\mathrm{th}}^{s}}$, where $N_{\mathrm{th}}^{s}$ is thermal noise.

\vspace{-5mm}
\subsection{Steady State \gls{snr} Estimation}
\label{subsec:snr_estimate}

Let us consider instantaneous downlink \gls{snr} as a random process where $\mu_{SNR^{S}}$ and $\sigma_{SNR^{S}}^2$ denote the mean and variance of the \gls{snr}, respectively. 
\gls{acm} implements a deterministic mapping from instantaneous \gls{snr} to a discrete \gls{mcs}. 
Let us define the ordered set of \gls{snr} thresholds as:
\begin{equation}
    ^{min}SNR^{S}_{1} <\cdots< SNR^{S}_{k} < \cdots < ^{max}SNR^{S}_{K}
\end{equation}
The system employs a \gls{mcs} $k$ with modulation order $M_k$ (e.g., QPSK, 16-QAM, 64-QAM) and the associated coding rate $r_k \in (0,1]$ if the $SNR^{S}$ falls between $k$ and $k-1$ \gls{snr} thresholds. 
The corresponding spectral efficiency is :
\begin{equation}
    \eta_k(SNR^{S}) = r_k \log_{2} M_k \quad \text{[bits/s/Hz]}.
\end{equation}
Thus, for system bandwidth $B^s$, the instantaneous capacity becomes:
\begin{equation}
C(t) = B^s \cdot \eta_k(SNR^{S}(t)) \quad [\text{bits/s}],
\label{eq:capacity}
\end{equation}

In Starlink's downlink operation, high \gls{snr} periods enable higher-order modulation, yielding increased spectral efficiency $\eta$ and consequently higher capacity $C(t)$. Conversely, \gls{snr} degradation due to increased slant range, atmospheric effects, or handovers forces \gls{acm} to select lower modulation orders, substantially reducing spectral efficiency. This creates a propagation chain where \gls{snr} variance directly influences spectral efficiency variance, and ultimately capacity variance. 
% While $\mathbb{E}[C]$ reflects average throughput, $\mathrm{Var}[C]$ determines service stability and predictability, a critical \gls{qos} factor. 
% Higher $\mathrm{Var}[C]$ results in pronounced throughput fluctuations that degrade user experience. Thus, the \gls{snr} distribution fundamentally governs both average capacity and its temporal stability.

\vspace{-5mm}
\color{black}
\subsection{Packet Drop Probability Due to Capacity Limitation}
\label{subsec:capacity_drop}

The time-varying Starlink capacity in Eq.~\eqref{eq:capacity} determines the instantaneous service rate available to the packet stream. 
For a link without buffering, the instantaneous capacity-induced packet drop probability can be approximated as the fraction of traffic that cannot be served:
\begin{equation}
P_{\mathrm{drop}}^{\mathrm{cap}}(t)
=
\left[
1 - \frac{C(t)}{R_{\mathrm{in}}(t)}
\right]^+,
\label{eq:instant_capacity_drop}
\end{equation}
where $[x]^+ = \max(x,0)$ and $R_{\mathrm{in}}$ is the corresponding traffic demand. 
Thus, no packets are dropped when $C(t)\geq R_{\mathrm{in}}(t)$, while drops occur when the instantaneous Starlink capacity falls below the offered traffic rate.

For a finite-buffer system, let $Q(t)$ denote the queue occupancy in packets and let $Q_{\max}$ be the buffer capacity. 
The queue evolves according to
\begin{equation}
Q(t+\Delta t)
=
\min
\left[
Q_{\max},
\left[
Q(t) + A(t,\Delta t) - S(t,\Delta t)
\right]^+
\right],
\end{equation}
where $A(t,\Delta t)$ is the number of arriving packets during interval $\Delta t$, and
\begin{equation}
S(t,\Delta t)
=
\left\lfloor
\frac{C(t)\Delta t}{\ell_{\mathrm{p}}}
\right\rfloor
\end{equation}
is the number of packets that can be served by the link, where $\ell_{\mathrm{p}}$ is the average packet size. 
The number of dropped packets in the interval is therefore
\begin{equation}
D(t,\Delta t)
=
\left[
Q(t) + A(t,\Delta t) - S(t,\Delta t) - Q_{\max}
\right]^+.
\end{equation}
Hence, the empirical packet drop probability due to capacity limitation is
\begin{equation}
P_{\mathrm{drop}}^{\mathrm{cap}}
=
\frac{\sum_{t} D(t,\Delta t)}
{\sum_{t} A(t,\Delta t)}.
\label{eq:empirical_capacity_drop}
\end{equation}

Since $C(t)$ is determined by the \gls{snr}-dependent \gls{mcs} selection, the capacity-limited drop probability can also be expressed over the discrete \gls{mcs} states. 
Let $C_k = B^s r_k\log_2 M_k$ be the capacity under \gls{mcs} state $k$, and let
\begin{equation}
p_k =
\Pr\left(
SNR^{S}_{k-1} \leq SNR^{S}(t) < SNR^{S}_{k}
\right)
\end{equation}
be the probability that the channel operates under state $k$. 
For a constant offered traffic rate $R_{\mathrm{in}}$, the average capacity-induced drop probability can be approximated as
\begin{equation}
\bar{P}_{\mathrm{drop}}^{\mathrm{cap}}
=
\sum_{k=1}^{K}
p_k
\left[
1 - \frac{C_k}{R_{\mathrm{in}}}
\right]^+.
\label{eq:mcs_capacity_drop}
\end{equation}
% This formulation directly links the \gls{snr} distribution, \gls{acm} decisions, spectral efficiency, and packet-level reliability. 
Therefore, high-\gls{snr} states yield larger $C_k$ and thus reduce the probability of capacity-induced packet loss, whereas low-\gls{snr} states caused by increased slant range, atmospheric attenuation, or handover-related degradation reduce $C_k$ and increase the likelihood of queue buildup and packet drops.
\color{black}

\vspace{-5mm}
\subsection{Probability of Failure due to Atmospheric Attenuation}
\label{subsec:Atmospheric_fail}

The \gls{itu} proposes a comprehensive estimation model for total attenuation ($A_T$) incorporating rain, cloud, gaseous, and tropospheric scintillation effects in satellite-to-ground links. This combined attenuation is expressed as~\cite{ITU‑R‑P.618‑14}:
\begin{equation}
\hspace{-3pt}
A_T(p) =
\begin{cases}
A_G(p) + \sqrt{(A_R(p)\!+\!A_C(p))^2 + A_S^2(p)}, \\ & \hspace{-50pt} : 0.001\%\le p \le 5\%, \\[1pt]
A_G(p) + \sqrt{A_C^2(p) + A_S^2(p)}, \\ & \hspace{-50pt} : 5\% < p \le 50\%,
\end{cases}
\label{eq:attenuation_threshold}
\end{equation}
where $p$ represents the probability that attenuation exceeds the specified threshold, and $A_R$, $A_C$, $A_G$, and $A_S$ denote rain, cloud, gaseous, and scintillation attenuation components, respectively.

To accurately model link availability across varying elevation angles, we define $\theta_{\min}$ as the minimum usable elevation angle for \gls{leo}-to-ground connections and partition the operational elevation range into bins $B_i = [\theta_i, \theta_{i+1})$ (typically 5° increments as recommended in \cite{ITU‑R‑P.618‑14}). Given a time horizon $T>0$ and instantaneous satellite elevation $\theta(t)$, the temporal fraction that the link spends in bin $B_i$ is:
\begin{equation}
  w_i
  \;=\;
  \frac{1}{T}\int_0^T
  \mathbb{I}\!\left\{\theta(t)\ge \theta_{\min}\right\}\,
  \mathbb{I}\!\left\{\theta_i \le \theta(t) < \theta_{i+1}\right\}\,dt,
\end{equation}
For a specified attenuation threshold $M$ (dB), we define $p_i(M)$ as the percentage of time within bin $B_i$ during which total atmospheric attenuation exceeds $M$, where $A_T^{(\theta_i)}(p_i) = M$. The overall probability of link failure due to excessive atmospheric attenuation is therefore:
\begin{equation}
  p^{\text{at}}(M) \;\approx\; \sum_i w_i\,p_i(M).
  \label{eq:at_prob}
\end{equation}
This elevation-weighted approach accounts for the nonuniform distribution of satellite positions and the corresponding variation in atmospheric path lengths.

\vspace{-5mm}
\subsection{Inter Satellite Links Analysis}
\label{subsec:isl}

\begin{figure}[!htb]
\centering
\includegraphics[width=0.9\linewidth]{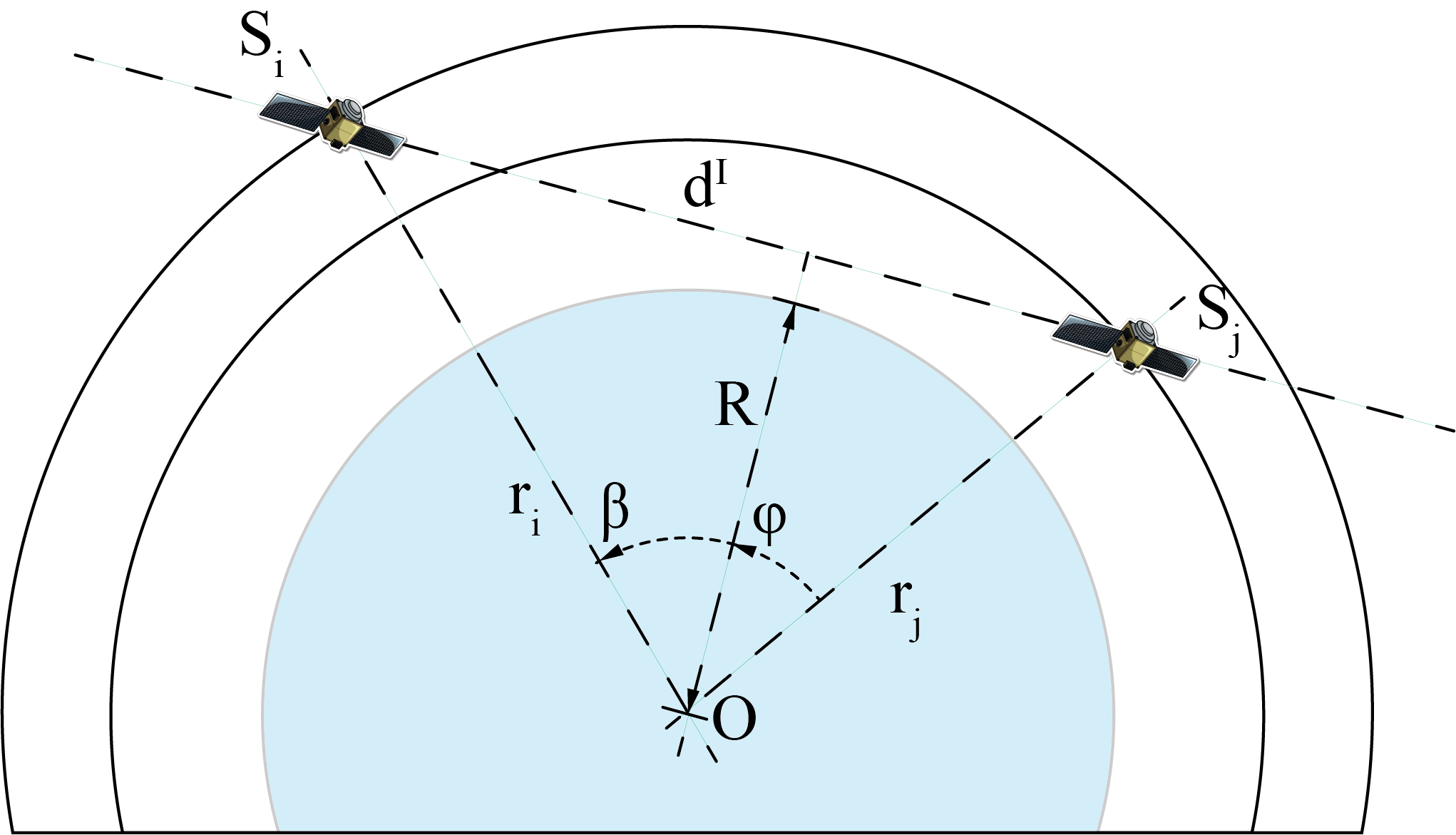}
\caption{ISLs between different orbits.}
\label{fig:isl}
\end{figure}

\begin{figure*}[!htb]
\centering
    \includegraphics[width=\textwidth]{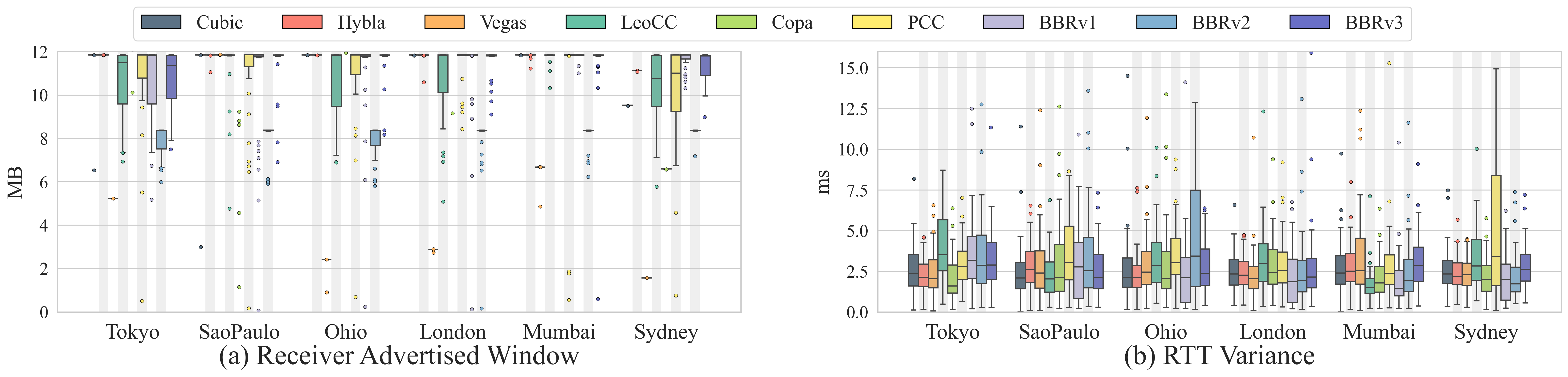}
  \caption{Receiver advertised window and \gls{rtt} variance of \glspl{cca} in dedicated downlink.}
\label{fig.download_seq_fig2_rwnd_rttvar}
\end{figure*}

\begin{figure*}[!htb]
\centering
    \includegraphics[width=\textwidth]{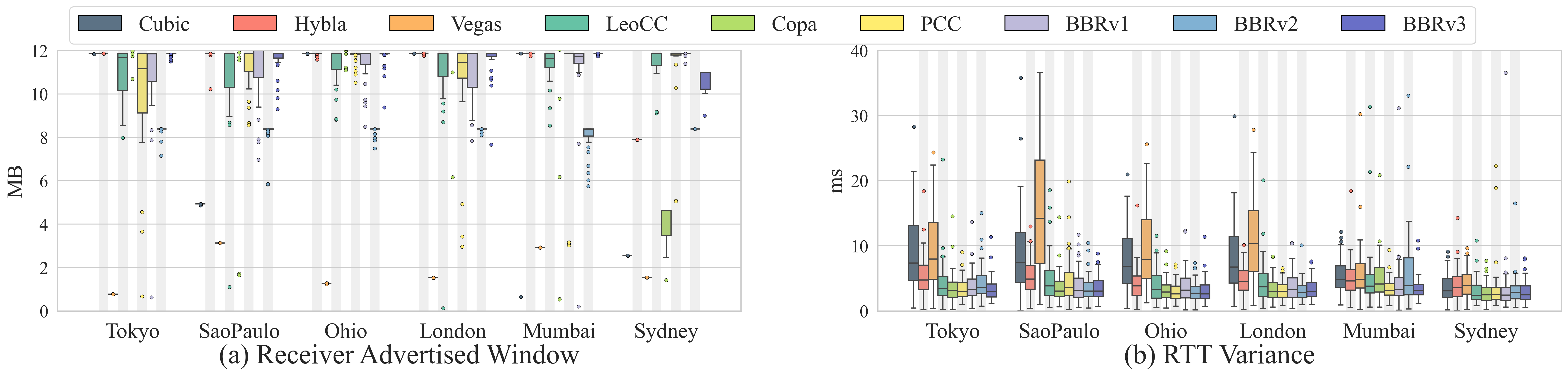}
  \caption{Receiver advertised window and \gls{rtt} variance of \glspl{cca} in concurrent downlink.}
\label{fig.download_competitive_fig2_rwnd_rttvar}
\end{figure*}

\begin{figure*}[!htb]
\centering
    \includegraphics[width=\textwidth]{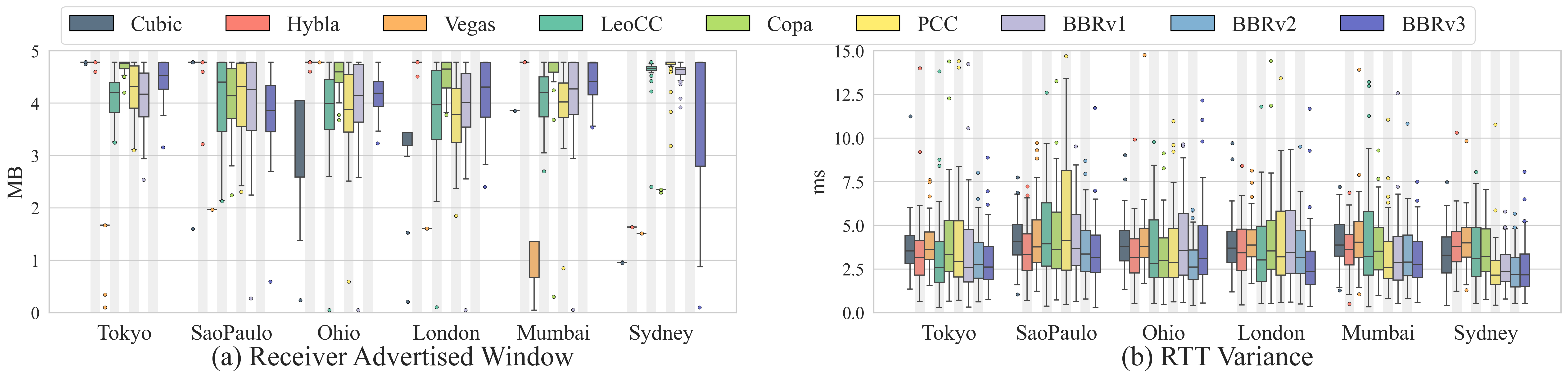}
  \caption{Receiver advertised window and \gls{rtt} variance of \glspl{cca} in dedicated uplink.}
\label{fig.uplink_seq_fig2_rwnd_rttvar}
\end{figure*}

\begin{figure*}[!htb]
\centering
    \includegraphics[width=\textwidth]{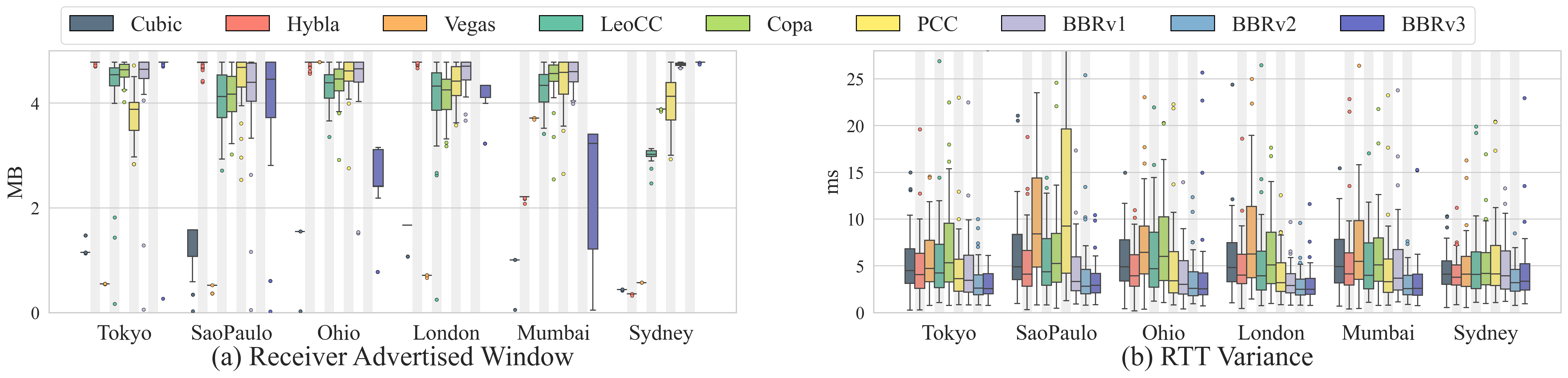}
  \caption{Receiver advertised window and \gls{rtt} variance of \glspl{cca} in concurrent uplink.}
\label{fig.uplink_competitive_fig2_rwnd_rttvar}
\end{figure*}

\begin{figure*}[!htb]
\centering
    \includegraphics[width=0.9\textwidth]{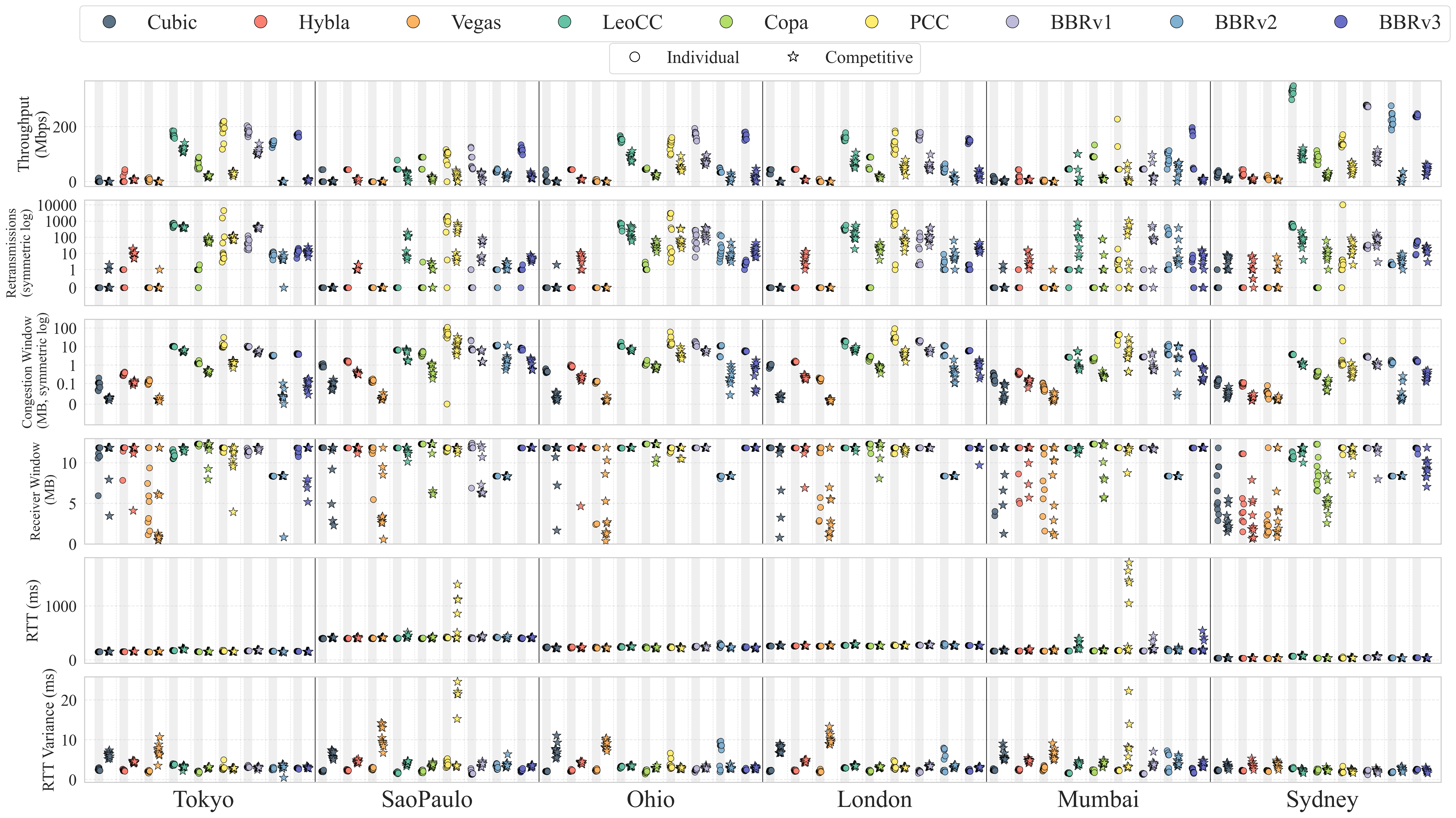}
      \caption{\textcolor{black}{Median downlink performance of CCAs across 10 experimental runs under individual and concurrent conditions. Each marker corresponds to a single experimental run across 10 repetitions. Circles indicate individual flows and stars indicate concurrent flows.}}
\label{fig:downlink_10_runs}
\end{figure*}

\begin{figure*}[!h]
\centering
    \includegraphics[width=\textwidth]{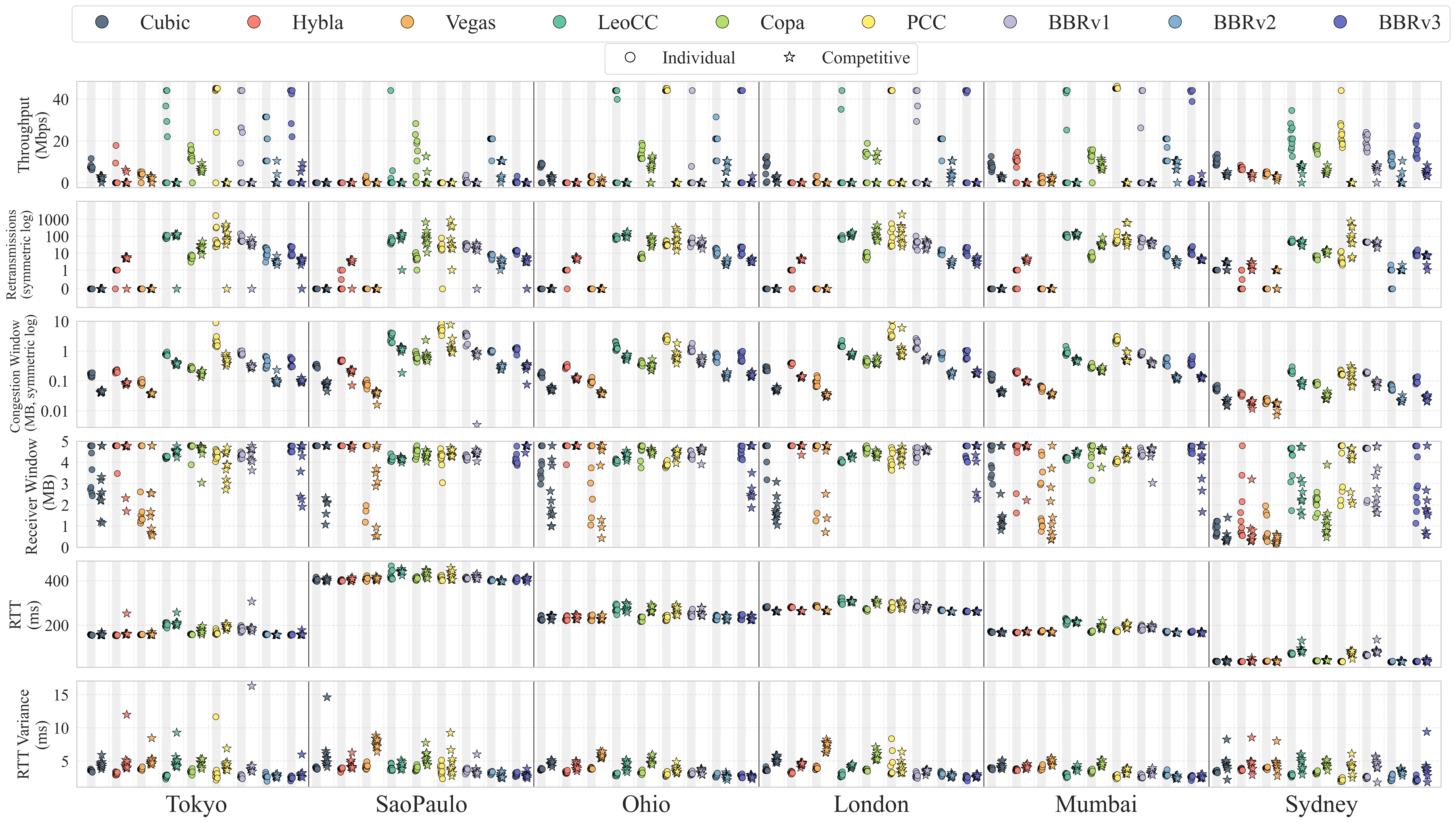}
  \caption{\textcolor{black}{Median uplink performance of CCAs across 10 experimental runs under individual and concurrent conditions. Each marker corresponds to a single experimental run across 10 repetitions. Circles indicate individual flows and stars indicate concurrent flows.}}
\label{fig:uplink_10_runs}
\end{figure*}

\begin{figure*}[!h]
\centering
    \includegraphics[width=0.9\textwidth]{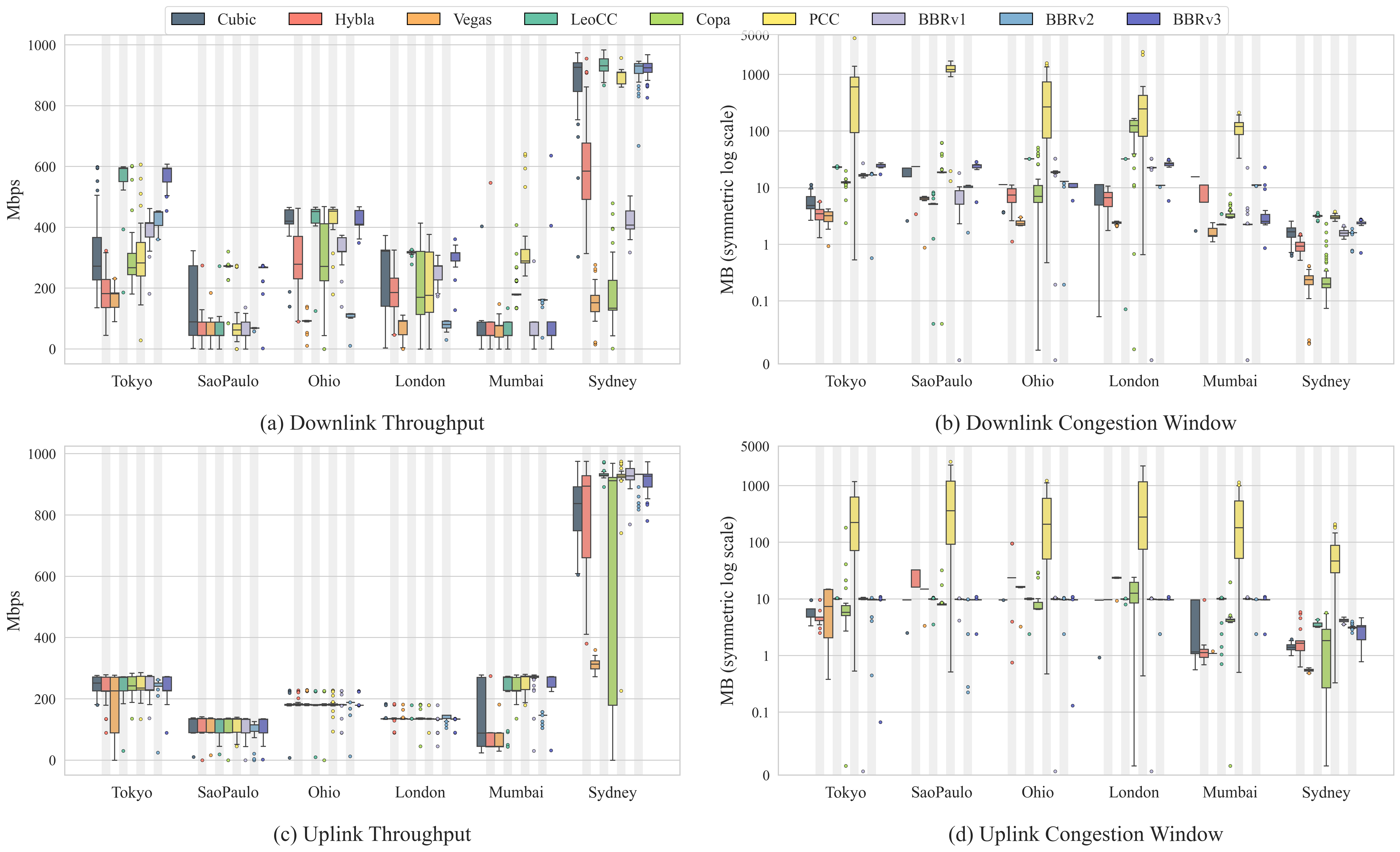}
  \caption{\textcolor{black}{Dedicated downlink and uplink performance over terrestrial network for the globally distributed server locations with different CCAs}}
\label{fig.terres_performance}
\end{figure*}

We now develop a model for optical inter-satellite links between satellites in different orbital planes. Consider two satellites $S_i$ and $S_j$ as depicted in Fig. \ref{fig:isl}, with optical wavelength $\lambda$ and telescope apertures $D_t$ and $D_r$ for transmitter and receiver, respectively.
The free space path loss of the optical \gls{isl} link between $S_i$ and $S_j$ can be defined as:
\begin{equation}
L_{\mathrm{fs}}^{I}(t) \;=\; \left(\frac{4\pi d^I(t)}{\lambda}\right)^{\!2}
\label{eq:fspl}
\end{equation}
where $d^I$ is the signal traveling distance. 
\begin{equation}
d^I(t) = \sqrt{r_i^2+r_j^2-2r_ir_j\cos\beta(t)}
\label{eq:dij}
\end{equation}
\begin{equation}
\beta = \arccos\left[\sin(\phi_i) \sin(\phi_j) + \cos(\phi_i) \cos(\phi_j) \cos(\Psi_i - \Psi_j)\right]
\end{equation}
where $(\phi_i, \Psi_i)$ and $(\phi_j, \Psi_j)$ refers to latitude and longitude position information of $S_i$ and $S_j$ satellites, respectively \cite{zhu2022laser}. 
Further, $r_i=R+h_i$ and $r_j=R+h_j$, where $h_i$ and $h_j$ are orbital altitudes of the two satellites of interest as detailed in Fig~\ref{fig:isl}. 
For a continuous transmission through the \gls{isl} without interruptions, it should uphold $r_j \cos (\phi) > R$ and $r_i \cos (\beta-\phi) > R$. 
In a given \gls{isl}, for transmit power $P_t$, the received optical power is:
\begin{equation}
P_r^{I}(t) \;=\; \frac{P_t^{I} \,\eta_t^{I} \,\eta_r^{I} \, G_t ^{I}\, G_r^{I}}{L_{\mathrm{fs}}^{I}(t)\,}
\label{eq:Pr_linear}
\end{equation}
where $\eta_t^{I} ,\eta_r^{I}$ are efficiencies, and $G_t ^{I}, G_r^{I}$ are gains of transmitter and receiver \gls{isl} antennas receptively. 
% With system noise temperature $T_s$ and noise figure $F$, the thermal noise can be defined as $N_{\mathrm{th}} \;=\; k\,T_s\,B\,F$ where $k$ is Boltzmann’s constant.
Thus, the \gls{isl} capacity is given as:
\begin{equation}
C_{\mathrm{ISL}}(t)
\;=\;
B^{I} \,\log_2\!\left(1+\frac{P_r(t)}{N_{\mathrm{th}}}\right).
\label{eq:isl_capacity_final}
\end{equation}
where $B^{I}$ is the bandwidth of the optical \gls{isl} and $N_{\mathrm{th}}$ is the thermal noise. 
In the Starlink constellation, co-orbital and adjacent-plane neighbors are permanent (availability $\approx 100\%$ over 24\,h), while some non-adjacent planes are intermittent with multiple disconnections per day \cite{zhu2022laser}.
Since each satellite is equipped with only a limited number of laser terminals (typically 3)~\cite{update-starlink}, the aggregate node capacity is constrained by the sum of the effective \gls{isl} capacities of its terminals.

If we assume that all the \glspl{isl} in the constellation have equal probability of failure, and the link failure and the recovery are modeled as Poisson processes,  the probability of an \gls{isl} failure can be given as \cite{song2024analysis}: 
\begin{equation}
    p^{\text{isl}} = \frac{\lambda_{on}^{\text{isl}}}{\lambda_{on}^{\text{isl}} + \lambda_{off}^{\text{isl}}} 
\end{equation}
where $\lambda_{on}^{\text{isl}}$ is the \gls{isl} failure arrival rate or the rate at which a functional \gls{isl} transitions into the disrupted state, and $\lambda_{off}^{\text{isl}}$ is the recovery arrival rate or the rate at which a disrupted link transitions back into the functional state.

\vspace{-2mm}
\color{black}
\subsection{Impact of Handover}
\label{subsec:handover}

Handovers in the Starlink constellation can be characterized as a schedule-driven process, subject to a system-wide synchronization \cite{hreha2019synchronization, starlink_fcc}. 
The handover timing is determined in advance by the gateway processor, leveraging satellite location, beam pattern, beam-hopping plan, and user terminal position. 
Based on the information, the gateway broadcasts handover data containing the terminal identity, source beam, target beam, and handover time. 
The terminal stores this information and performs the transition at the scheduled instant by returning to the new beam and updating its beam-hopping state. 
Near the handover instant, the system may also operate with wider \gls{acm}/\gls{tdm} margins to improve robustness. 
Thus, a handover failure can arise primarily due to: (i) failure to receive or decode the broadcast handover command correctly, (ii) timing misalignment among the satellite, gateway, and user terminal, and (iii) terminal retuning or switching failure during execution, assuming there is no failure to provision or activate the target beam and associated beam-hopping state at the scheduled epoch. 
% This view is consistent with the patent description, which emphasizes coordinated gateway-driven handover and system-wide synchronization using a master clock, timing beacons, and gateway-to-terminal timing data. 
Therefore, the per-attempt handover failure probability of user terminal $j$ at slot $n$ can be expressed as:
\begin{equation}
p_j^{\mathrm{ho}}[n]
=
1-
\big(1-p_{\mathrm{ctrl},j}[n]\big)
\big(1-p_{\mathrm{sync},j}[n]\big)
\big(1-p_{\mathrm{sw},j}[n]\big),
\label{eq:scheduled_ho_fail}
\end{equation}
where $p_{\mathrm{ctrl},j}[n]$ is the probability of handover-control decoding failure, $p_{\mathrm{sync},j}[n]$ is the probability of synchronization failure, and $p_{\mathrm{sw},j}[n]$ is the probability of switching failure at the terminal.

\color{black}

Therefore, the probability of link failure causing packet loss in the transmission path ($p^{\text{tot}}$) can be modeled as:
\begin{equation}
    p^{\text{tot}} = 1 - (1-p^{cap}) \cdot(1-p^{gw}) \cdot (1-p^{at})\cdot (1-p^{ho}) \cdot \prod_{i=1}^{N^{\text{isl}}} (1-p_i^{\text{isl}} )
    \label{eq:total_prop1_2}
\end{equation}
where $N^{\text{isl}}$ is the number of \glspl{isl} that a given packet is routed through in the constellation, $p^{gw} = p^{at}(M')$ (Eq. \eqref{eq:at_prob}), and the other parameters represent the respective failure probabilities~\cite{latency2023pan}.

\color{black}
\section{Extensive Evaluation}
\label{appendixC}

Fig.~\ref{fig.download_seq_fig2_rwnd_rttvar} and Fig.~\ref{fig.download_competitive_fig2_rwnd_rttvar} present the \gls{cca} receiver-advertised window and \gls{rtt} variance of dedicated and concurrent downlinks, respectively. 
On the other hand, Fig.~\ref{fig.uplink_seq_fig2_rwnd_rttvar} and Fig.~\ref{fig.uplink_competitive_fig2_rwnd_rttvar} illustrate the \gls{cca} receiver-advertised window and \gls{rtt} variance of dedicated and concurrent in the given order. 
To explore the consistency of the testbed results, we repeat all four test cases 10 times over the Starlink Internet. 
Fig.~\ref{fig:downlink_10_runs} depicts the medians of the six key measured performance indicators, both dedicated and concurrent downlinks, for the \glspl{cca} of interest. In addition, Fig.~\ref{fig:uplink_10_runs} presents the medians of the same key parameters of dedicated and concurrent uplink measurements. 
Furthermore, in order to compare the Starlink Internet performance with terrestrial networks, we repeat dedicated uplink and downlink tests over the University network. 
The same testbed was connected to the University network, and the measured throughput and congestion window are presented in Fig~\ref{fig.terres_performance}(a) and (b) for downlink and Fig~\ref{fig.terres_performance}(c) and (d) for uplink.
The University network throughput is limited to 1~Gbps at the switch port, thus underscoring the maximum experienced throughput in both uplink and downlink. 
These throughput results further confirm that the Starlink transmission path is the primary bottleneck in our distributed testbed, not \gls{aws} network constraints. 

\color{black}

\textbf{\begin{figure}[!htb]
\centering
\includegraphics[width=\linewidth]{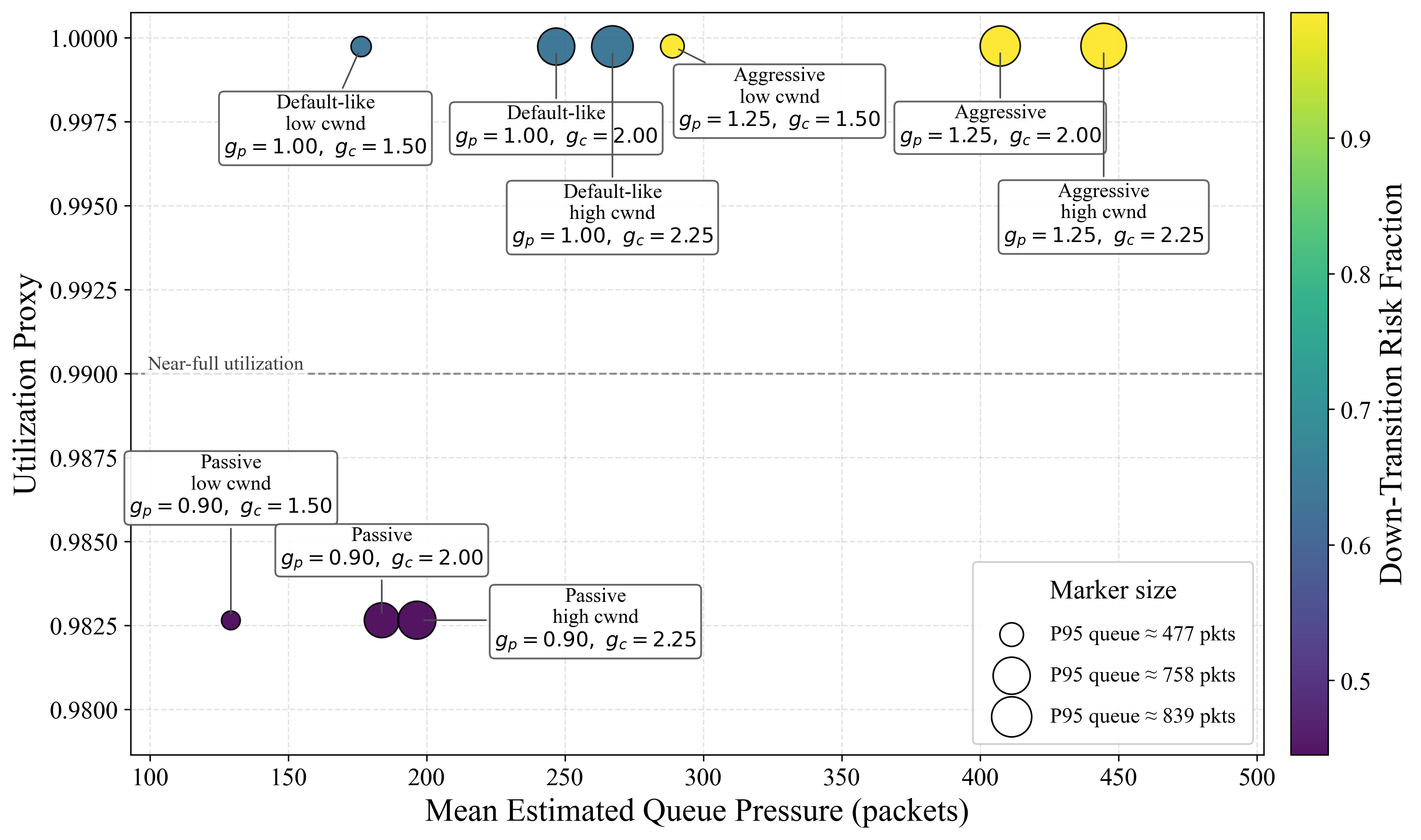}
\caption{\textcolor{black}{Trace-Driven BBR-v3 gain sensitivity analysis for Starlink Internet}}
\label{fig:parameter_sensitivity}
\end{figure}}

% \begin{figure*}[!t]
% \centering
%     \includegraphics[width=0.9\textwidth]{Images/log-scale/terrestrial_uplink_downlink_throughput_cwnd_2x2_log_cwnd_revised_copa_pcc.png}
%     \caption{Dedicated downlink and uplink performance over terrestrial network 
%     for the globally distributed server locations with different CCAs}
%     \label{fig.terres_performance}

%     \vspace{2mm}

%     \begin{minipage}{0.48\textwidth}
%     \centering
%     \includegraphics[width=\linewidth]{Images/sydney_downlink_bbrv3_gain_sensitivity.png}
%     \caption{Trace-Driven BBR-v3 gain sensitivity analysis for Starlink Internet}
%     \label{fig:parameter_sensitivity}
%     \end{minipage}
% \end{figure*}

\color{black}

\section{Sensitivity Analysis of \gls{bbrv3} Parameters}
\label{appendixD}

% The live measurement campaign uses the default Linux \gls{bbrv3} configuration to ensure reproducibility, deployability, and a fair comparison with real end-host operation. 
In order to examine whether Starlink-specific parameter tuning could improve \gls{bbrv3}, we perform a trace-driven sensitivity analysis using the proposed fluid model. 
Let $g_p$ and $g_c$ denote pacing-gain and congestion-window-gain perturbation factors, respectively. 
These are not intended to reproduce all internal Linux \gls{bbrv3} state-dependent constants; rather, they represent controlled variations around the default operating point. 
The pacing-gain range $g_p\in\{0.9,1.0,1.25\}$ follows the drain, cruise, and probe intuition of BBR's ProbeBW behavior.
The congestion window gain range $g_c\in\{1.5,2.0,2.5\}$ varies around the BDP scaling values. 
Using the measured delivery rate ($\widehat{B}(t)$) and measured \gls{rtt} ($\text{RTT}(t)$) from the Sydney dedicated downlink traces, we estimate the time-varying path state as:
\begin{equation}
\widehat{\mathrm{BDP}}(t)=\widehat{B}(t)\text{RTT}_{\min}(t),
\end{equation}
where $\text{RTT}_{\min}(t)$ is the rolling minimum \gls{rtt}. 
The gain-dependent pacing rate and target inflight volume can be modeled as:
\begin{equation}
r_{\mathrm{pace}}(t;g_p)=g_p\widehat{B}(t),
\end{equation}
\begin{equation}
v_{\mathrm{target}}(t;g_p,g_c)
=
\min\left(g_c\widehat{\mathrm{BDP}}(t),\,r_{\mathrm{pace}}(t;g_p)\text{RTT}(t)\right).
\end{equation}
The modeled inflight volume follows the same first-order relaxation principle used in the proposed fluid model, thus connecting the parameter sweep to the measured Starlink throughput and \gls{rtt} dynamics.
\begin{equation}
\frac{d v_{\mathrm{model}}(t;g_p,g_c)}{dt}
=
\frac{1}{T_s}
\left(
v_{\mathrm{target}}(t;g_p,g_c)
-
v_{\mathrm{model}}(t;g_p,g_c)
\right)
\end{equation}

\vspace{-1mm}
For each gain pair, we evaluate utilization, queue pressure, and retransmission-sensitive down-transition risk. The estimated queue pressure is computed as the excess modeled inflight volume above the measured BDP.
\begin{equation}
Q(t;g_p,g_c)
=
\left[
v_{\mathrm{model}}(t;g_p,g_c)-\widehat{\mathrm{BDP}}(t)
\right]^+
\end{equation}
while the down-transition risk is approximated by the fraction of samples where the modeled inflight exceeds the probing threshold or where the estimated queue becomes large:
\begin{multline}
R_{\mathrm{dwn}}(g_p,g_c) = \frac{1}{T} \sum_t
\mathbb{I}\Bigl(
v_{\mathrm{model}}(t;g_p,g_c) > 1.25\widehat{\mathrm{BDP}}(t) \\
\lor\; Q(t;g_p,g_c) > Q_{\mathrm{thr}}
\Bigr).
\end{multline}
The results illustrated in Fig.~\ref{fig:parameter_sensitivity} show a clear Starlink-specific tuning trade-off. 
Conservative pacing, e.g., $g_p=0.9$, reduces queue pressure and down-transition risk, but it underutilizes the measured downlink for a significant fraction of the trace. 
Increasing $g_p$ and $g_c$ moves the model toward near-full utilization. 
When utilization saturates, additional aggressiveness mainly increases $Q(t)$ and $R_{\mathrm{dwn}}$ rather than producing meaningful throughput gain. 
This indicates that fixed aggressive gains are not well matched to Starlink's non-stationary capacity, \gls{rtt}-shift, and handover-driven dynamics.
Therefore, an improved QoS could be delivered through Starlink-oriented \gls{bbrv3} configuration, reducing aggressiveness during high \gls{rtt} variance, retransmission bursts, or queue-buildup intervals, and restoring higher gains only during stable high-capacity periods.

\color{black}

% \vspace{-2mm}
% \bibliographystyle{IEEEtran} %BST
% \bibliography{mybib}

% \end{document}

\end{document}